\documentclass[11pt,a4paper]{article}
\usepackage[onehalfspacing]{setspace}

\usepackage[english]{babel}

\usepackage[]{placeins}

\usepackage{etex}

\usepackage{amsfonts}
\usepackage{natbib}											
\usepackage{graphicx, adjustbox} 									
\usepackage[small,bf,hang]{caption}			
\usepackage{subcaption}									
\usepackage{wrapfig}										
\usepackage{multirow}										
\usepackage{amsmath}										
\allowdisplaybreaks[1]									
\usepackage{amssymb}										
\usepackage{amsthm}											
\usepackage{booktabs}										
\usepackage{colortbl}
\usepackage{url}                        
\usepackage{eurosym}                    
\usepackage{textcomp}                   
\usepackage{fancyhdr}                   
\usepackage{listings} 									
\usepackage{pdfpages}										
\usepackage{float}                      
\usepackage[title,titletoc,toc]{appendix}				
\usepackage{titling}

\usepackage{enumerate}									
\usepackage[multiple]{footmisc}																	
\usepackage{enumitem}										
\usepackage{mathtools}										
\usepackage{IEEEtrantools}								
\usepackage{standalone}	
\usepackage{algorithm}										
\usepackage{algorithmic}									
\usepackage{authblk}											
\usepackage{tikz}												
\usetikzlibrary{patterns}
\usepackage[group-separator={\,}]{siunitx}

\pgfdeclarepatternformonly{thick horizontal lines}{\pgfqpoint{-2pt}{0pt}}{\pgfqpoint{10pt}{10pt}}{\pgfqpoint{10pt}{10pt}}%
{
  \pgfsetlinewidth{2pt}
  \pgfpathmoveto{\pgfqpoint{-1pt}{0pt}}
  \pgfpathlineto{\pgfqpoint{9pt}{0pt}}
  \pgfusepath{stroke}
}

\pgfdeclarepatternformonly{thick vertical lines}{\pgfqpoint{0pt}{-2pt}}{\pgfqpoint{10pt}{10pt}}{\pgfqpoint{10pt}{10pt}}%
{
  \pgfsetlinewidth{2pt}
  \pgfpathmoveto{\pgfqpoint{0pt}{-1pt}}
  \pgfpathlineto{\pgfqpoint{0pt}{9pt}}
  \pgfusepath{stroke}
}

\usepackage[pdftex,colorlinks, citecolor=blue, urlcolor=blue,linkcolor=red]{hyperref}

\usepackage{bm}

\usepackage{grffile}
\usepackage{amsfonts}
\usepackage{pdflscape}
\usepackage{amssymb}
\usepackage{adjustbox}
\usepackage{rotating}
\usepackage{siunitx}
\usepackage{cprotect}
\usepackage{color}	
\usepackage{etoolbox}

\urldef{\CBSlinkdxt}\url{http://statline.cbs.nl/Statweb/publication/?DM=SLNL&PA=37530ned&D1=0%2c2%2c4&D2=1-100&D3=0&D4=60-63&HDR=T%2cG1%2cG2&STB=G3&VW=T}

\urldef{\CBSlinkpxt}\url{http://statline.cbs.nl/Statweb/publication/?DM=SLNL&PA=7461BEV&D1=0&D2=a&D3=1-100&D4=60-63&HDR=T%2cG3&STB=G1%2cG2&VW=T}

\makeatletter

\pretocmd{\NAT@citex}{%
  \let\NAT@hyper@\NAT@hyper@citex
  \def\NAT@postnote{#2}%
  \setcounter{NAT@total@cites}{0}%
  \setcounter{NAT@count@cites}{0}%
  \forcsvlist{\stepcounter{NAT@total@cites}\@gobble}{#3}}{}{}
\newcounter{NAT@total@cites}
\newcounter{NAT@count@cites}
\def\NAT@postnote{}

\def\NAT@hyper@citex#1{%
  \stepcounter{NAT@count@cites}%
  \hyper@natlinkstart{\@citeb\@extra@b@citeb}#1%
  \ifnumequal{\value{NAT@count@cites}}{\value{NAT@total@cites}}
    {\ifNAT@swa\else\if*\NAT@postnote*\else%
     \NAT@cmt\NAT@postnote\global\def\NAT@postnote{}\fi\fi}{}%
  \ifNAT@swa\else\if\relax\NAT@date\relax
  \else\NAT@@close\global\let\NAT@nm\@empty\fi\fi
  \hyper@natlinkend}
\renewcommand\hyper@natlinkbreak[2]{#1}

\patchcmd{\NAT@citex}
  {\ifNAT@swa\else\if*#2*\else\NAT@cmt#2\fi
   \if\relax\NAT@date\relax\else\NAT@@close\fi\fi}{}{}{}
\patchcmd{\NAT@citex}
  {\if\relax\NAT@date\relax\NAT@def@citea\else\NAT@def@citea@close\fi}
  {\if\relax\NAT@date\relax\NAT@def@citea\else\NAT@def@citea@space\fi}{}{}

\makeatother

\definecolor{darkgreen}{RGB}{40,150,40}

\lstset{
 language=R,                
 basicstyle=\scriptsize,       
 numbers=left,                   
 numberstyle=\scriptsize,      
 stepnumber=1,                   
 numbersep=5pt,                  
 keywordstyle=\color{blue},
 identifierstyle=, 
 commentstyle=\color{darkgreen}, 
 stringstyle=\ttfamily, 
 backgroundcolor=\color{white},  
 showspaces=false,               
 showstringspaces=false,         
 showtabs=false,                 
 frame=single,                   
 tabsize=2,                      
 breaklines=true,                
 breakatwhitespace=false,        
 alsoletter={.},							
 otherkeywords={!,!=,~,$,*,\&,\%/\%,\%*\%,\%\%,<-,<<-,/,$},
 morekeywords={break, else, for, if, in, next, repeat, return, switch, try, while, array, character, complex, data.frame, double, function, integer, list, logical, matrix, numeric, vector, Inf, NA, NaN, NULL, FALSE, TRUE, set.seed}, 
 deletekeywords={\_, replace, start, data, quantile, factor, scale, old, exp, new, lower, upper, pt, col, c, order}, 
}
\lstset{literate=%
   *{0}{{{\color{blue}0}}}1
    {1}{{{\color{blue}1}}}1
    {2}{{{\color{blue}2}}}1
    {3}{{{\color{blue}3}}}1
    {4}{{{\color{blue}4}}}1
    {5}{{{\color{blue}5}}}1
    {6}{{{\color{blue}6}}}1
    {7}{{{\color{blue}7}}}1
    {8}{{{\color{blue}8}}}1
    {9}{{{\color{blue}9}}}1
}

\renewcommand{\leq}{\leqslant}					
\renewcommand{\geq}{\geqslant}					
\renewcommand{\epsilon}{\varepsilon}

\theoremstyle{definition}

\usepackage{empheq}
\usepackage[many]{tcolorbox}
\tcbset{
  myhlight/.style={
    colback=yellow,
    arc=0pt,
    outer arc=0pt,
    boxrule=0pt,
    top=2pt,
    bottom=2pt,
    left=2pt,
    right=2pt,
  },
  highlight math style={myhlight}
}

\newcommand{\revision}[1]{#1}

\def\IBNR{\mathrm{IBNR}}
\def\Rep{\mathrm{R}}
\def\Observed{\mathrm{Obs}}
\def\Hidden{\mathrm{Hidden}}

\interfootnotelinepenalty=10000

\DeclareCaptionLabelFormat{andtable}{#1~#2  \&  \tablename~\thetable}

\usepackage{graphicx}
\usepackage{subcaption}
\usepackage{lscape}
\usepackage{multirow}
\usepackage{setspace}
\usepackage[latin1]{inputenc}
\usepackage[a4paper, left=2.5cm, right=2.5cm, top=2cm, bottom=2cm, includehead, includefoot, head=30pt]{geometry}
\usepackage{amssymb}
\usepackage{amsmath}
\usepackage{amsthm}
\usepackage{mathrsfs}
\usepackage{booktabs}
\usepackage{url}
\usepackage{eurosym}
\usepackage{textcomp}
\usepackage{fancyhdr}
\usepackage{listings}
\usepackage{pdfpages}
\usepackage{float}
\usepackage[title,titletoc,toc]{appendix}
\usepackage{enumerate}
\usepackage{enumitem}
\usepackage{authblk}
\usepackage{tikz}
\usepackage{color}
\usepackage{etoolbox}
\usepackage[normalem]{ulem}
\usepackage{natbib}
\usepackage{diagbox}
\usepackage{environ}

\makeatletter
\newsavebox{\measure@tikzpicture}
\NewEnviron{scaletikzpicturetowidth}[1]{%
  \def\tikz@width{#1}%
  \begin{lrbox}{\measure@tikzpicture}%
  \BODY
  \end{lrbox}%
  \pgfmathparse{#1/\wd\measure@tikzpicture}%
  \BODY
}
\makeatother

\setcounter{MaxMatrixCols}{10}

\parindent 0pt
\parskip \medskipamount
\usetikzlibrary{arrows,decorations.pathmorphing,backgrounds,positioning,fit,petri}
\newtheorem*{example*}{Example}
\makeatletter
\bibpunct{(}{)}{;}{a}{,}{,}
   \setlength{\bibsep}{0pt}
\bibliographystyle{apaCustom}
\pagestyle{fancy}
\fancyhf{}

\fancyhf[HL]{\nouppercase{\textit{\leftmark}}}
\fancyhead[HR]{\thepage}

\newtheorem*{exampleSP*}{Example where the surrender value is part of the premiums paid}
\newtheorem*{examplePR*}{Example where the surrender value is part of the reserve}

\parskip=\medskipamount

\newcommand{\vectorF}[1]{\boldsymbol{#1}}
\newcommand{\characteristic}[1]{\mathbbm{1}_{#1}}

\usepackage{pgfplots,relsize}
\pgfplotsset{compat=1.14}
\usetikzlibrary{patterns}
\usepackage{bbm}
\usepackage{soul}
\usepackage{pdflscape}

\definecolor{graphBlue}{HTML}{56B4E9}
\definecolor{graphBlue1}{HTML}{56B4E9}
\definecolor{graphBlue2}{HTML}{6EBEEC}
\definecolor{graphBlue3}{HTML}{86C9EF}
\definecolor{graphBlue4}{HTML}{9ED4F2}
\definecolor{graphBlue5}{HTML}{B6DEF5}
\definecolor{graphBlue6}{HTML}{CEE9F8}
\definecolor{graphBlue7}{HTML}{E6F4FB}

\newcommand{\alternative}[1]{}
\usepackage{array}
\newcolumntype{K}[1]{>{\centering\arraybackslash}p{#1}}

\newcommand{\weight}[0]{exposure}
\newcommand{\weights}[0]{exposures}

\usepackage{tikz,ifthen,fp,calc}

\makeatletter
\newlength{\tkz@size}
\newlength{\tkz@rect@A}
\newlength{\tkz@rect@B}
\newlength{\tkz@rect@C}
\newlength{\tkz@rect@D}
\newlength{\tkz@hachsep}
\newboolean{tkz@rect@inv}\setboolean{tkz@rect@inv}{false}

\def\tkzhachrect[#1](#2,#3)(#4,#5){%

\draw (#2,#3) rectangle (#4,#5) ;
\setboolean{tkz@rect@inv}{false}
\setlength{\tkz@hachsep}{#1 cm}
\setlength{\tkz@rect@A}{#2 cm + #3 cm}
\setlength{\tkz@rect@B}{#2 cm + #5 cm}
\setlength{\tkz@rect@C}{#4 cm + #3 cm}

\ifthenelse{\lengthtest{\tkz@rect@B > \tkz@rect@C}}%
{\setlength{\tkz@rect@C}{#2 cm + #5 cm}
\setlength{\tkz@rect@B}{#4 cm + #3 cm}
\setboolean{tkz@rect@inv}{true}%
}{}%
\setlength{\tkz@rect@D}{#4 cm + #5 cm}
\setlength{\tkz@size}{\tkz@rect@A}

\whiledo{\lengthtest{\tkz@size < \tkz@rect@D}}%
{\ifthenelse{\lengthtest{\tkz@size < \tkz@rect@B}}
    {\draw[hstyle] (#2 cm,\tkz@size-#2 cm) -- (\tkz@size-#3 cm,#3 cm);}
    {\ifthenelse{\lengthtest{\tkz@size < \tkz@rect@C}}
       {\ifthenelse{\boolean{tkz@rect@inv}}
       {\draw[hstyle] (#2 cm,\tkz@size-#2 cm) -- (#4 cm,\tkz@size-#4 cm);}
       {\draw[hstyle] (\tkz@size - #5 cm,#5 cm) -- (\tkz@size-#3 cm,#3 cm);}%
       }%
    {\draw[hstyle] (\tkz@size - #5 cm,#5 cm) -- (#4 cm,\tkz@size-#4 cm);}}
    \addtolength{\tkz@size}{\tkz@hachsep}
}
}

\def\tkzhachrectfp[#1](#2,#3)(#4,#5){%
\setboolean{tkz@rect@inv}{false}
\FPadd{\deb}{#2}{#3}
\FPtrunc\deb{\deb}{2}
\FPadd{\fin}{#4}{#5}
\FPtrunc\fin{\fin}{2} 
\FPadd{\sone}{#2}{#5}
\FPtrunc\sone{\sone}{2}
\FPadd{\stwo}{#4}{#3}
\FPtrunc\stwo{\stwo}{2} 
\FPifgt{\sone}{\stwo}
\FPset{\temp}{\sone}
\FPset{\sone}{\stwo}
\FPset{\stwo}{\temp}
\setboolean{tkz@rect@inv}{true}%
\else
\fi
\FPadd{\hach}{\deb}{#1}%
\FPtrunc\hach{\hach}{2}%

\draw (#2,#3) rectangle (#4,#5);

\foreach \s in {\deb ,\hach,...,\sone}
   {\FPadd{\oo}{\s}{-#2} 
    \FPtrunc\oo{\oo}{2}%
    \FPadd{\aa}{\s}{-#3} 
    \FPtrunc\aa{\aa}{2}%
    \draw[hstyle] (#2,\oo) -- (\aa,#3);}%
 \FPifeq{\sone}{\stwo}%
\else
   \FPadd{\sone}{\sone}{#1}
   \FPadd{\hach}{\sone}{#1}
   \FPtrunc\hach{\hach}{2}
     \foreach \s in {\sone ,\hach,...,\stwo}
    {\ifthenelse{\boolean{tkz@rect@inv}}
       {\FPadd{\oo}{\s}{-#2} 
       \FPtrunc\oo{\oo}{2}
        \FPadd{\aa}{\s}{-#4} 
        \FPtrunc\aa{\aa}{2}
        \draw[hstyle] (#2,\oo) -- (#4,\aa);}
       {\FPadd{\oo}{\s}{-#5} 
       \FPtrunc\oo{\oo}{2}
        \FPadd{\aa}{\s}{-#3} 
        \FPtrunc\aa{\aa}{2}
        \draw[hstyle] (\oo,#5) -- (\aa,#3);}%
    }
\fi%

\FPadd{\stwo}{\stwo}{#1}
\FPadd{\hach}{\stwo}{#1}
\FPtrunc\hach{\hach}{2}

 \foreach \s in {\stwo,\hach,...,\fin}
 {\FPadd{\oo}{\s}{-#5} \FPtrunc\oo{\oo}{2}
  \FPadd{\aa}{\s}{-#4} \FPtrunc\aa{\aa}{2}
    \draw[hstyle] (\oo,#5) -- (#4,\aa);}
 }

\usepackage{mathtools}
\usepackage{graphicx}
\usepackage{environ}

\NewEnviron{resizealign}{\sbox0{
    $\begin{matrix}\displaystyle\BODY\end{matrix}$}%
  \sbox1{$(\theequation)$}%
  \sbox2{\parbox{\dimexpr \wd0 + 2\wd1}%
    {\begin{align}\BODY\end{align}}}
  \noindent\resizebox{\textwidth}{!}{\usebox2}%
}

\graphicspath{{Images/}}

\newcommand{\partialD}[2]{\frac{\partial #1}{\partial #2}}
\newcommand{\partialDD}[3]{\frac{\partial #1}{\partial #2 \partial #3}}

\usepackage[latin1]{inputenc}
\usepackage[T1]{fontenc}
\usepackage{lmodern}

\newcommand{\covariateDisplay}[2]{#1}

\makeatletter
\def\today{%
  \ifcase\month\or%
  January\or February\or March\or April\or May\or June\or%
  July\or August\or September\or October\or November\or December\fi \,%
  \two@digits{\the\day} , 
  \number\year%
}
\makeatother

\usepackage{csquotes}
\DeclareQuoteStyle{english}
  {\textquotedblleft}
  [\textquotedblleft]
  {\textquotedblright}
  [0.05em]
  {\textquoteleft}
  {\textquoteright}
\MakeOuterQuote{"}

\begin{document}
\title{Modeling the number of hidden events\\subject to observation delay}
\author[1,3,*]{Jonas Crevecoeur}
\author[1,2,3,4]{Katrien Antonio}
\author[1,3,4]{Roel Verbelen}
\affil[1]{Faculty of Economics and Business, KU Leuven, Belgium.}
\affil[2]{Faculty of Economics and Business, University of Amsterdam, The Netherlands.}
\affil[3]{LRisk, Leuven Research Center on Insurance and Financial Risk Analysis, KU Leuven, Belgium.}
\affil[4]{LStat, Leuven Statistics Research Center, KU Leuven, Belgium.}
\affil[*]{Corresponding author. E-mail: jonas.crevecoeur@kuleuven.be}
\date{\today}
\maketitle

\begin{abstract}
\noindent Copyright \copyright\ 2019 European Journal of Operational Research. This paper considers the problem of predicting the number of events that have occurred in the past, but which are not yet observed due to a delay. Such delayed events are relevant in predicting the future cost of warranties, pricing maintenance contracts, determining the number of unreported claims in insurance and in modeling the outbreak of diseases. Disregarding these unobserved events results in a systematic underestimation of the event occurrence process. Our approach puts emphasis on modeling the time between the occurrence and observation of the event, the so-called observation delay. We propose a granular model for the heterogeneity in this observation delay based on the occurrence day of the event and on calendar day effects in the observation process, such as weekday and holiday effects. We illustrate this approach on a European general liability insurance data set where the occurrence of an accident is reported to the insurer with delay.

\end{abstract}


\section{Introduction}

\begin{figure}[ht!]
\center
\begin{tikzpicture}[scale=0.7, auto, to/.style={->,>=stealth',shorten >=1pt}, every node/.style={font=\fontsize{9pt}{9pt}\selectfont\sffamily, align=center, semithick}]

\draw[to] (0,0) -- (20,0) node[below right]  {};
\draw[to] (0, 0) -- (0, 8) node[above] {Time since start};

\draw[<->, ,>=stealth] (0,-1.6) --  node[below]  {Observation window} (13,-1.6);
\draw[<->, ,>=stealth] (13,-1.6) -- node[below]  {Future} (20,-1.6);


\draw (1.5, 0) -- (10.5, 6);

\node[circle,draw, inner sep=.3cm,
           path picture={
               \node at (path picture bounding box.center){
                   \hspace{0.10cm}\includegraphics[width=.8cm]{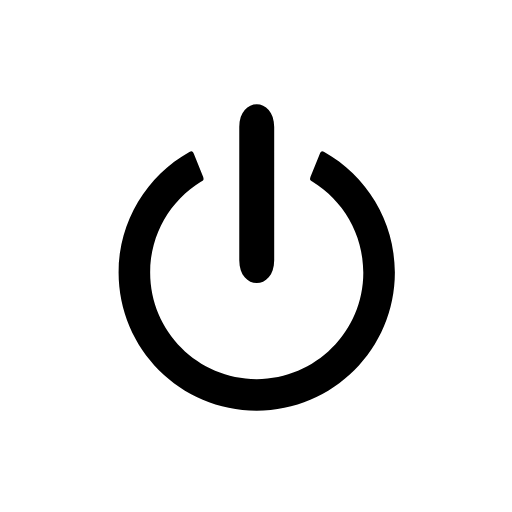}
               };
           },
			fill=white] at (1.5, 0) {};

\node at (1.5, 1) {Start};
\node at (1.5, -1) {1};

\node[circle,draw, inner sep=.3cm,
           path picture={
               \node at (path picture bounding box.center){
                   \hspace{0.10cm}\includegraphics[width=1cm]{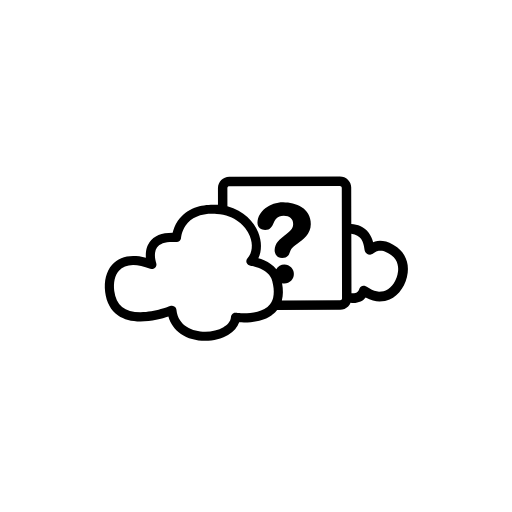}
               };
           },
			fill=white] at (6.75, 3.5) {};

\node at (6.75, 4.7) {Hidden\\event};

\node[circle,draw, inner sep=.3cm,
           path picture={
               \node at (path picture bounding box.center){
                   \hspace{0.10cm}\includegraphics[width=.7cm]{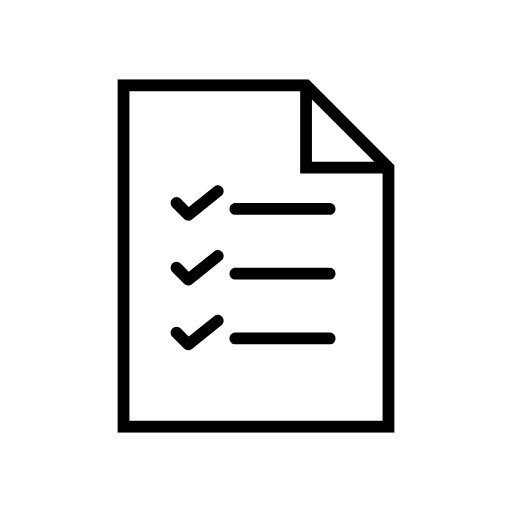}
               };
           },
			fill=white] at (10.5, 6) {};

\node at (10.5, 7.2) {Observed \\ event};

tikz={\draw [decorate, line width=1pt, decoration={brace, mirror, amplitude=3mm}, sloped, anchor=center] (2.55, 0.3) -- (6.3, 2.8) node [midway, below=4mm] {Event delay};}
tikz={\draw [decorate, line width=1pt, decoration={brace, mirror, amplitude=3mm}, sloped, anchor=center] (7.65, 3.7) -- (10.05, 5.3) node [midway, below=4mm] {Observation delay};}


\draw (6.5, 0) -- (13, 4.33);
\draw[dashed] (13, 4.33) -- (16.25, 6.5);

\node[circle,draw, inner sep=.3cm,
           path picture={
               \node at (path picture bounding box.center){
                   \hspace{0.10cm}\includegraphics[width=.8cm]{iconStart.png}
               };
           },
			fill=white] at (6.5, 0) {};
\node at (6.5, -1) {2};

\node[circle,draw, inner sep=.3cm,
           path picture={
               \node at (path picture bounding box.center){
                   \hspace{0.10cm}\includegraphics[width=1cm]{iconHidden.png}
               };
           },
			fill=white] at (9.5, 2) {};

\node[circle,draw, inner sep=.3cm,
           path picture={
               \node at (path picture bounding box.center){
                   \hspace{0.10cm}\includegraphics[width=.7cm]{iconReport.png}
               };
           },
			fill=white] at (16.25, 6.5) {};


\draw (8.5, 0) -- (13, 3);
\draw[dashed] (13, 3) -- (18.25, 6.5);

\node[circle,draw, inner sep=.3cm,
           path picture={
               \node at (path picture bounding box.center){
                   \hspace{0.10cm}\includegraphics[width=.8cm]{iconStart.png}
               };
           },
			fill=white] at (8.5, 0) {};
\node at (8.5, -1) {3};


\draw (10.5, 0) -- (13, 1.67);
\draw[dashed] (13, 1.67) -- (18.75, 5.5);

\node[circle,draw, inner sep=.3cm,
           path picture={
               \node at (path picture bounding box.center){
                   \hspace{0.10cm}\includegraphics[width=.8cm]{iconStart.png}
               };
           },
			fill=white] at (10.5, 0) {};
\node at (10.5, -1) {4};

\node[circle,draw, inner sep=.3cm,
           path picture={
               \node at (path picture bounding box.center){
                   \hspace{0.10cm}\includegraphics[width=1cm]{iconHidden.png}
               };
           },
			fill=white] at (15.75, 3.5) {};


\draw[thick] (13, 0) -- (13, 8) node [above] {Evaluation date};

\end{tikzpicture}

\caption{Occurrence and observation of events}
\label{figure:developmentScenario}
\end{figure}

In many domains within operational research analysts are interested in building a stochastic model for the occurrence of events. However, the events of interest are often observed or reported with some delay. Analysts should account for these unobserved events since ignoring them will bias the decisions based on the stochastic model under consideration. Figure~\ref{figure:developmentScenario} visualizes this setting. We specify a well defined observation window (on the $x$-axis) in which we observe the creation of new objects (e.g.~products or contracts). Over the course of their lifetimes some objects may experience the event of interest (object 1 and 2 in Figure~\ref{figure:developmentScenario}) before a given evaluation date, and others will not (object 3 and 4 in Figure~\ref{figure:developmentScenario}). Upon occurrence the event is initially hidden from the decision maker. The time that elapses between the onset of the object's lifetime and the occurrence of the event is called the event delay. Only after a so-called observation or reporting delay the decision maker becomes aware of the existence of the event.  This paper outlines a data driven strategy to predict the number of events that occurred in the past (before the evaluation date), but which are hidden at the time of evaluation and will only be observed or reported in the future. Subject 2 in Figure~\ref{figure:developmentScenario} is an example of such an event.

The modeling of the time to occurrence of an event (`the event delay'), the number of (hidden) events that occurred during a specific time window and the delay between occurrence and observation (`the observation delay') have been active research areas in the literature on operational research, actuarial science and epidemiology. Typical examples of applications where this predictive problem matters are: a portfolio of maintenance, warranty or insurance contracts, but also an outbreak of a specific disease fits within this framework. We highlight some relevant contributions and explain how this paper extends the existing literature. 

A warranty contract requires the manufacturer to compensate the buyer for all failures occurring within the warranty period.  Manufacturers hold capital for future compensations related to goods produced in the past. The amount of capital required depends on the number of defective products that have been sold. Accurate estimation of this number is complicated due to the incompleteness of the data. The diagonal time line in Figure~\ref{figure:developmentScenario} begins when a defective product is produced. However, the warranty period only starts when the product is sold to a customer. Manufacturers are typically not aware of these sales and we consider them as a hidden events. Once the defect emerges and the customer calls his warranty contract, the manufacturer is informed of the sale (`the observed event'). \cite{Akbarov2012} and \cite{ye2014} simultaneously model the time to sale and the delay between sale and failure of the product using parametric methods. Since both processes interact in the likelihood, estimation is difficult. \cite{Akbarov2012} resolve to numerical maximization, whereas \cite{ye2014} use a Stochastic Expectation Maximization strategy. While these authors model the time to sale with a simple, parametric distribution without covariates, our framework accounts for the seasonal effects, promotions holidays and weather effects typically present in sales data. 

Epidemiologists face a similar statistical problem when modeling the evolution of diseases \citep{Harris1990, Salmon2015}. In this setting, subjects are followed over time and a recent disease infection may remain unobserved due to either delay in disease diagnosis by a medical doctor or incubation time. Modeling these delays allows to take the yet unobserved infections (`the hidden events') into account and thus enables a faster and more accurate identification of disease outbreaks and epidemics \citep{Noufaily2016}.

Maintenance contracts are typically sold together with large industrial appliances. Under these contracts the manufacturer or a third party guarantees the continued use of the equipment. A machine failure (`the observed event') is often the result of previous defects (`the hidden event') which remained unobserved. These defects can be detected by on site inspections and timely repairs will prevent expensive failures or breakdowns of the machine. However, the profitability of these inspections depends largely on the number of hidden defects. Observation delay was first modelled in the context of maintenance contracts by \cite{Christer1973}, where it is called delay-time. Since then several papers have focussed on the delay-time concept. \cite{Baker1993} model delay-time from observed failure data using maximum likelihood estimation. In this approach both the time to defect as well as the time to observation of the machine failure are tackled with parametric distributions. This literature typically assumes a constant intensity for the occurrences of defects and ignores heterogeneity in the delay-time distribution. \cite{Wang1997} and \cite{Apeland2003} rely on expert opinions to formulate a fully subjective delay-time model. \cite{Wang2010} and \cite{Berrade2018} focus on economic decision making when the delay-time distribution is known. In line with the current era of big data analytics (see \cite{mortenson2015}), our approach goes beyond these assumptions and proposes a data driven strategy to capture heterogeneity in both the occurrence of defects as well as in the delay between a defect and its observation. 


The case-study presented in this paper illustrates our data driven approach with an insurance data set where contracts are sold to policyholders. Some policyholders will be involved in an accident or other type of insured event, while others will not. In insurance parlance the delay between the occurrence (`the hidden event') of an accident and the reporting or filing of the claim to the insurance company (`the observed event') is called the reporting delay. These delays are strongly portfolio dependent and can be substantial when the insured does not immediately notice the damage. In the remainder of the paper we only consider accidents that will eventually be reported. Accidents that are never reported do not get reimbursed and are not relevant for the balance sheet of the insurer. Once the claim is reported and accepted by the insurer, the insurer reimburses the loss with a single payment or a series of payments. Insurance companies book a reserve to be able to settle the claims that are Incurred But Not yet Reported (IBNR) and refer to this capital as the IBNR reserve. Estimating the number of claims from past exposures that will be reported beyond the evaluation date (the so-called IBNR claim counts) is crucial in setting this reserve. Motivated by computational constraints from the past, many estimation methods in insurance structure the data from Figure~\ref{figure:developmentScenario} in a two dimensional table that aggregates the number of accidents by their year of occurrence and year of reporting. We refer the reader to \cite{Taylor2000, WuthrichMerz2008, WuthrichMerz2015} for more details on reserving with aggregate methods. Relatively few papers address the problem of specifying a model at granular level for the phenomenon sketched in Figure~\ref{figure:developmentScenario}. \cite{badescu2016marked} and \cite{avanzi2016micro} focus on modeling the accident arrival process at a weekly level using Cox processes. These models allow to capture over-dispersion and serial dependence, which is often encountered in such occurrence data. The assumption of independence between the occurrence date and the reporting delay is a disadvantage of the models presented in \cite{badescu2016marked} and \cite{avanzi2016micro}. \cite{Verrall2016} were the first to present a model for IBNR counts at a daily level, including the heterogeneity in reporting delays based on the occurrence date of the claim and the strong weekday pattern leading to less claims being reported during the weekend. This weekday pattern relates to calendar day effects in the reporting process which are difficult to model using classical techniques designed for aggregated data (see \cite{Kuang2008}).
\cite{Verrall2016} provide a method to incorporate this weekday pattern for reporting delays of less than one week. \cite{Verbelen2017} extend this weekday pattern to reporting delays beyond the first week by separately estimating weekly and intra week reporting probabilities. Moreover, \cite{Verbelen2017} present the Expectation Maximization algorithm as a framework for jointly estimating the occurrence and reporting process.

Our paper models the occurrence of hidden events non-parametrically. This allows to capture fluctuations in occurrence counts (for example due to seasonality or weather conditions) without explicitly modeling these events. Moreover, extending the work of \cite{Verrall2016} and \cite{Verbelen2017} we model the observation delay in the presence of multiple covariates, including calendar day effects. Examples of such calendar day effects are: a reduction in observed events during the weekend, the effect of national holidays and seasonality in observation delay. Our strategy introduces the concept of observation exposure as an intuitive and flexible framework for incorporating (multiple) calendar day effects through regression. This approach elegantly transforms the observation delay distribution by scaling the probability of observing an event on a certain date based on covariates. As such, the transformed observation delay distribution becomes independent of these covariates and is then modelled with a simple, parametric distribution.  This makes our approach suitable to a wide range of problems.

This paper is organized as follows. Section~\ref{section:model} describes a statistical framework for modeling the number of hidden events subject to an observation delay. In Section~\ref{section:caseStudy} we illustrate this approach in a case-study involving an insurance data set. We also investigates the performance of our model in four simulated scenarios. The online appendix provides detailed expressions for implementing the model and links our approach to the non-parametric Kaplan-Meier estimator \citep{KaplanMeier}.

\section{A granular model for the occurrence of events subject to delay} \label{section:model}
Denote by $N_t$ the number of events occurring on date $t$, where $t = 1$ is the date of the first event. These events remain hidden until their observation at date $s$ after a delay $s-t$. Let $N_{t, s}$ be the number of events that occurred on date $t$ and are observed on date $s$. Since all events will be observed at some point in the future, we find
\begin{equation*}
	N_{t} = \sum_{s \geq t} N_{t, s}.
\end{equation*}
Consider an evaluation date $\tau$ at which we have to predict the number of hidden events. At $\tau$ we split the events from a past occurrence date $t$ into observed ($s \leq \tau$) and hidden events which are not yet observed ($s > \tau$), respectively denoted by
\begin{equation*}
	N_t^{\Observed}(\tau) = \sum_{s = t}^{\tau} N_{t, s} \quad \text{and} \quad N_t^{\Hidden}(\tau) = \sum_{s = \tau + 1}^{\infty} N_{t, s} \quad \text{for } t \leq \tau.
\end{equation*}
We obtain the total number of hidden events by aggregating the unobserved events from all past occurrence dates, i.e.
\begin{equation*}
	N^{\Hidden}(\tau) = \sum_{t = 1}^\tau N_t^{\Hidden}(\tau) = \sum_{t = 1}^\tau \sum_{s = \tau + 1}^{\infty} N_{t, s}.
\end{equation*}
This total count is the number that we want to predict. Following \cite{Jewell1990} and \cite{Norberg1993}, we formulate two distributional assumptions from which the number of hidden events can be predicted:
\begin{itemize}
	\item[(A1)] The event occurrence process $(N_t)_{t \geq 1}$ follows an inhomogeneous Poisson \revision{distribution} with intensity $(\lambda_t)_{t \geq 1}$.
	\item[(A2)] \revision{The observation delay is independent and identically distributed for events occurring on the same date.}
\end{itemize}
Denote by $p_{t, s}$ the probability of observing an event from occurrence date $t$ on date $s$. We use the notation $p_{t}^{\Observed}(\tau)$ for the probability that an event from date $t$ is observed by the evaluation date $\tau$. This probability is
\begin{equation*}
	p_t^{\Observed}(\tau) = \sum_{s = t}^\tau p_{t, s}.
\end{equation*}
By assumption (A1) and (A2) the conditions for the Poisson thinning property \citep{PoissonProcesses} are satisfied. The thinning property implies that all $N_{t, s}$ are independent and
\begin{equation}
	N_{t, s} \sim \text{Poisson}(\lambda_t \cdot p_{t, s}). \label{eq:poissonThinning}
\end{equation}
This allows us to construct the likelihood for the observed data at time $\tau$. Let $\vectorF{\chi}$ denote the available data, consisting of all events that are observed on the evaluation date $\tau$
\begin{equation*}
	\vectorF{\chi} = \{N_{t, s} \mid t \leq s \leq \tau \}.
\end{equation*}
The loglikelihood of the observed data is
\begin{equation}
	\ell(\vectorF{\lambda}, \vectorF{p} ; \vectorF{\chi} ) = \sum_{t = 1}^\tau \sum_{s = t}^\tau \Big[N_{t, s} \cdot \log(\lambda_t) + N_{t, s} \cdot \log ( p_{t, s}) - \lambda_t \cdot p_{t, s} - \log(N_{t, s}!) \Big] \label{equation:loglikelihood}
\end{equation}
where $\vectorF{\lambda}$ is a vector with components $\lambda_t$ for observed occurrence dates $t$ and $\vectorF{p} = \{p_{t, s} \, \mid \, t \leq s \leq \tau \}$. This paper puts focus on the observation process without imposing any structure on $\lambda_t$. A straightforward computation shows that the loglikelihood in $\eqref{equation:loglikelihood}$ is maximal for
\begin{equation}
	\lambda_t = \frac{\sum_{s=t}^{\tau} N_{t, s}}{\sum_{s=t}^\tau p_{t, s}} = \frac{N_t^{\Observed}(\tau)}{p_t^{\Observed}(\tau)}. \label{eq:zeta}
\end{equation}
Replacing $\lambda_t$ by this expression the loglikelihood in $\eqref{equation:loglikelihood}$ becomes
\begin{equation}
 	\ell(\vectorF{p} ; \vectorF{\chi} ) = \sum_{t = 1}^\tau \sum_{s = t}^\tau N_{t, s} \cdot \log(p_{t, s}) - \sum_{t=1}^\tau N_t^{\Observed}(\tau) \cdot \log(p_t^{\Observed}(\tau))  + \text{constants}. \label{eq:likelihoodgeneric}
\end{equation}
Up to constants this is the loglikelihood for a right truncated observation delay random variable. The truncation point is $\tau - t$, which is the maximal observed delay for an event that occurred on date $t$.

\subsection{A time change strategy to model observation delay} \label{section:dailyReportingWeights}

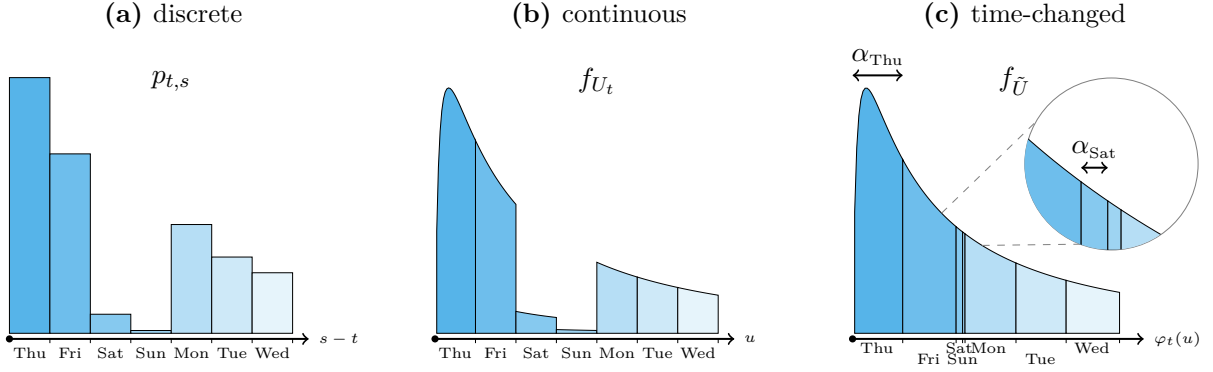
\begin{figure}
\center
\begin{subfigure}[t]{.30\textwidth}
\caption{discrete}\label{figure:timeChangeDiscrete}
\begin{tikzpicture} [scale=0.95]
	\begin{axis}[width=1.45\linewidth, hide axis, xmin=-0.2, xmax=2.2, ymax=1.02, ymin=-0.11, at={(0*\textwidth, 0)}]
    \addplot[domain=0:0.25, fill=graphBlue1] {0.912}\closedcycle;
    \addplot[domain=0.25:0.5, fill=graphBlue2] {0.640}\closedcycle;
    \addplot[domain=0.5:0.75, fill=graphBlue3] {0.068}\closedcycle;
    \addplot[domain=0.75:1, fill=graphBlue4] {0.01}\closedcycle;
    \addplot[domain=1:1.25, fill=graphBlue5] {0.388}\closedcycle;
    \addplot[domain=1.25:1.5, fill=graphBlue6] {0.272}\closedcycle;
    \addplot[domain=1.5:1.75, fill=graphBlue7] {0.216}\closedcycle;
	
	\draw[<->, thick, opacity=0] (axis cs:0, 0.92) -- node[above, opacity=0]{$\alpha_{thu, 0}$} (axis cs: 0.31, 0.92) ;
	
	\draw[->, thick] (axis cs:0, -0.02) -- (axis cs: 1.85, -0.02) node[right]{\tiny $s-t$};

    \node[fill,circle, inner sep=1pt] at (axis cs:0, -0.02) {};

    \draw (axis cs:0.25,-0.02) -- (axis cs:0.25,-0.035);
    \draw (axis cs:0.5,-0.02) -- (axis cs:0.5,-0.035);
    \draw (axis cs:0.75,-0.02) -- (axis cs:0.75,-0.035);
    \draw (axis cs:1,-0.02) -- (axis cs:1,-0.035);
    \draw (axis cs:1.25,-0.02) -- (axis cs:1.25,-0.035);
    \draw (axis cs:1.5,-0.02) -- (axis cs:1.5,-0.035);
    \draw (axis cs:1.75,-0.02) -- (axis cs:1.75,-0.035);

    \node at (axis cs:0.125,-0.070) {\tiny Thu};
    \node at (axis cs:0.375,-0.07) {\tiny Fri};
    \node at (axis cs:0.625,-0.07) {\tiny Sat};
    \node at (axis cs:0.875,-0.07) {\tiny Sun};
    \node at (axis cs:1.125,-0.07) {\tiny Mon};
    \node at (axis cs:1.375,-0.07) {\tiny Tue};
    \node at (axis cs:1.625,-0.07) {\tiny Wed};

    \node at (axis cs:1, 0.90) {$p_{t, s}$};

    \end{axis}
\end{tikzpicture}
\end{subfigure}
\hfill
\begin{subfigure}[t]{.30\textwidth}
\caption{continuous}\label{figure:timeChangeContinuous}
\begin{tikzpicture} [scale=0.95]	
    \begin{axis}[width=1.45\linewidth, hide axis, xmin=-0.2, xmax=2.2, ymax=1.02, ymin=-0.11, at={(0, 0)}]
    \addplot[domain=0:0.25, fill=graphBlue1] {1/x/1.574961/sqrt(2*pi)*exp(-(ln(x))^2/2/2.48}\closedcycle;
	 \addplot[domain=0.25:0.50, fill=graphBlue2] {1/x/1.574961/sqrt(2*pi)*exp(-(ln(x))^2/2/2.48}\closedcycle;
	 \addplot[domain=0.50:0.75, fill=graphBlue3] {0.17*1/x/1.574961/sqrt(2*pi)*exp(-(ln(x))^2/2/2.48}\closedcycle;
	 \addplot[domain=0.75:1, fill=graphBlue4] {0.04*1/x/1.574961/sqrt(2*pi)*exp(-(ln(x))^2/2/2.48}\closedcycle;
	 \addplot[domain=1:1.25, fill=graphBlue5] {1/x/1.574961/sqrt(2*pi)*exp(-(ln(x))^2/2/2.48}\closedcycle;
	 \addplot[domain=1.25:1.50, fill=graphBlue6] {1/x/1.574961/sqrt(2*pi)*exp(-(ln(x))^2/2/2.48}\closedcycle;
     \addplot[domain=1.50:1.75, fill=graphBlue7] {1/x/1.574961/sqrt(2*pi)*exp(-(ln(x))^2/2/2.48}\closedcycle;

 	\draw[->, thick] (axis cs:0, -0.02) -- (axis cs: 1.85, -0.02) node[right]{\tiny $u$};

	\draw[<->, thick, opacity=0] (axis cs:0, 0.92) -- node[above, opacity=0]{$\alpha_{thu, 0}$} (axis cs: 0.31, 0.92) ;

     \node[fill,circle, inner sep=1pt] at (axis cs:0, -0.02) {};

    \draw (axis cs:0.25,-0.02) -- (axis cs:0.25,-0.035);
    \draw (axis cs:0.5,-0.02) -- (axis cs:0.5,-0.035);
    \draw (axis cs:0.75,-0.02) -- (axis cs:0.75,-0.035);
    \draw (axis cs:1,-0.02) -- (axis cs:1,-0.035);
    \draw (axis cs:1.25,-0.02) -- (axis cs:1.25,-0.035);
    \draw (axis cs:1.5,-0.02) -- (axis cs:1.5,-0.035);
    \draw (axis cs:1.75,-0.02) -- (axis cs:1.75,-0.035);

    \node at (axis cs:0.125,-0.070) {\tiny Thu};
    \node at (axis cs:0.375,-0.07) {\tiny Fri};
    \node at (axis cs:0.625,-0.07) {\tiny Sat};
    \node at (axis cs:0.875,-0.07) {\tiny Sun};
    \node at (axis cs:1.125,-0.07) {\tiny Mon};
    \node at (axis cs:1.375,-0.07) {\tiny Tue};
    \node at (axis cs:1.625,-0.07) {\tiny Wed};
	
	\node at (axis cs:1, 0.90) {$f_{U_t}$};
	
    \end{axis}

\end{tikzpicture}
\end{subfigure}
\hfill
\begin{subfigure}[t]{.30\textwidth}
\caption{time-changed}\label{figure:timeChangeFull}
\begin{tikzpicture} [scale=0.95]	
    \begin{axis}[width=1.45\linewidth, hide axis, xmin=-0.2, xmax=2.2, ymax=1.02, ymin=-0.11, at = {(0, 0)}]
    \addplot[domain=0:0.31, fill=graphBlue1] {1/x/1.574961/sqrt(2*pi)*exp(-(ln(x))^2/2/2.48}\closedcycle;
	 \addplot[domain=0.31:0.64, fill=graphBlue2] {1/x/1.574961/sqrt(2*pi)*exp(-(ln(x))^2/2/2.48}\closedcycle;
	 \addplot[domain=0.64:0.68, fill=graphBlue3] {1/x/1.574961/sqrt(2*pi)*exp(-(ln(x))^2/2/2.48}\closedcycle;
	 \addplot[domain=0.68:0.695, fill=graphBlue4] {1/x/1.574961/sqrt(2*pi)*exp(-(ln(x))^2/2/2.48}\closedcycle;
	 \addplot[domain=0.695:1.01, fill=graphBlue5] {1/x/1.574961/sqrt(2*pi)*exp(-(ln(x))^2/2/2.48}\closedcycle;
	 \addplot[domain=1.01:1.32, fill=graphBlue6] {1/x/1.574961/sqrt(2*pi)*exp(-(ln(x))^2/2/2.48}\closedcycle;
     \addplot[domain=1.32:1.65, fill=graphBlue7] {1/x/1.574961/sqrt(2*pi)*exp(-(ln(x))^2/2/2.48}\closedcycle;

 	\draw[->, thick] (axis cs:0, -0.02) -- (axis cs: 1.8, -0.02) node[right]{\tiny $\varphi_t(u)$};

     \node[fill,circle, inner sep=1pt] at (axis cs:0, -0.02) {};

     \draw[<->, thick] (axis cs:0, 0.92) -- node[above]{$\alpha_{\text{\tiny Thu}}$} (axis cs: 0.31, 0.92) ;

     \draw[-, thin, gray, dashed] (axis cs:0.80, 0.3135) -- (axis cs:1.70, 0.32);
     \draw[-, thin, gray, dashed] (axis cs:0.55, 0.4285) -- (axis cs:1.25, 0.82);
     \draw[thin, gray, fill=white] (axis cs:1.60, 0.604) circle (1.2cm);

     \def\xpos{1.30};
     \def\maxX{1.5}
     \def\ypos{1};
     \def\scale{5};
     \def\yscale{1}

    \draw (axis cs:0.31,-0.02) -- (axis cs:0.31,-0.035);
    \draw (axis cs:0.64,-0.02) -- (axis cs:0.64,-0.035);
    \draw (axis cs:0.68,-0.02) -- (axis cs:0.68,-0.035);
    \draw (axis cs:0.695,-0.02) -- (axis cs:0.695,-0.035);
    \draw (axis cs:1.01,-0.02) -- (axis cs:1.01,-0.035);
    \draw (axis cs:1.32,-0.02) -- (axis cs:1.32,-0.035);
    \draw (axis cs:1.65,-0.02) -- (axis cs:1.65,-0.035);

    \node at (axis cs:0.15,-0.050) {\tiny Thu};
    \node at (axis cs:0.47,-0.09) {\tiny Fri};
    \node at (axis cs:0.66,-0.050) {\tiny Sat};
    \node at (axis cs:0.68,-0.09) {\tiny Sun};
    \node at (axis cs:0.84,-0.050) {\tiny Mon};
	\node at (axis cs:1.16,-0.09) {\tiny Tue};
	\node at (axis cs:1.48,-0.050) {\tiny Wed};
	
	\node at (axis cs:1, 0.90) {$f_{\tilde{U}}$};

    \end{axis}

    \begin{axis}[width=.81\linewidth, hide axis, xmin=0, xmax=1, ymax=1, ymin=0, at={(0.58\linewidth, 1.59cm)}] 

 		\begin{scope}
     \clip (axis cs:0.54, 0.68) circle (1.2cm);

     \addplot[domain=0:0.36, fill=graphBlue2] {(1/(x*0.25+0.55)/1.574961/sqrt(2*pi)*exp(-(ln((x*0.25+0.55)))^2/2/2.48) - 0.313)*8}\closedcycle;
		\addplot[domain=0.36:0.52, fill=graphBlue3] {(1/(x*0.25+0.55)/1.574961/sqrt(2*pi)*exp(-(ln((x*0.25+0.55)))^2/2/2.48) - 0.313)*8}\closedcycle;
		\addplot[domain=0.52:0.60, fill=graphBlue4] {(1/(x*0.25+0.55)/1.574961/sqrt(2*pi)*exp(-(ln((x*0.25+0.55)))^2/2/2.48) - 0.313)*8}\closedcycle;
		\addplot[domain=0.60:1, fill=graphBlue5] {(1/(x*0.25+0.55)/1.574961/sqrt(2*pi)*exp(-(ln((x*0.25+0.55)))^2/2/2.48) - 0.313)*8}\closedcycle;
		
		\draw[<->, thick] (axis cs:0.36, 0.65) -- node[above]{$\alpha_{ \text{\tiny Sat}}$} (axis cs: 0.52, 0.65) ;

 		\end{scope}
    \end{axis}
\end{tikzpicture}
\end{subfigure}
\caption{Observation delay distribution for an event that occurred on a Thursday. We illustrate {\bf (a)} the discrete observation delay probabilities $p_{t, s}$, {\bf (b)} the density of the continuous observation delay distribution $U_t$ and {\bf (c)} the density of the time-changed observation delay distribution $\tilde{U}$.} \label{figure:timeChange}
\end{figure}

We are interested in structuring the observation probabilities $p_{t, s}$ based on covariates corresponding to the occurrence date $t$ and the reporting date $s$ of the event. The probabilistic nature of the data enforces the constraints
\begin{align}
	p_{t, s} \geq 0, \quad  \forall t, s \quad \text{and} \quad \sum_{s \geq t} p_{t, s} = 1, \quad  \forall t. \label{eq:constraint}
\end{align}
The proposed time change strategy transforms the reporting probabilities such that they can be linked with covariates while preserving these constraints. This transformation is depicted in Figure~\ref{figure:timeChange}, where we consider an event that occurred on a Thursday and for which observation is less likely during the weekend.

First, we view the discrete observation delay as a realization of a continuous random variable $U_t$ under interval censoring. This is graphically illustrated in Figure~\ref{figure:timeChangeDiscrete} (discrete setting) and \ref{figure:timeChangeContinuous} (continuous setting). Second, we define a time change operator $\varphi_{t}$ which assigns a positive length $\alpha_{t, s}$, called the observation exposure, to each combination of an occurrence date $t$ and an observation date $s$. This time change operator is similar to the concept of operational time, which is a common technique in continuous financial mathematics, see \cite{Swishchuk2016}. We perceive dates as having variable lengths, whereas prior to this time change an equal length of one time unit was attached to each date. The probability of observing an event on a certain date is scaled by the duration of this date, which motivates calling this length the observation exposure. We define the time-changed delay $\varphi_t(d)$ for an event with occurrence date $t$ and an observation delay of $d$ days as

\begin{equation}
	\varphi_{t}(0) = 0 \quad \text{and} \quad \varphi_{t}(d) = \sum_{i=1}^d \alpha_{t, t+i-1}, \quad d \in \mathbb{N} \setminus \{0\}. \label{eq:transformation}
\end{equation}
This is the sum of all observation exposures $\alpha_{t, s}$ assigned to dates in between the occurrence date $t$ and date $t+d-1$. By applying $\varphi_t$ on the observation delay random variable $U_t$ we obtain a time-changed random variable $\tilde{U} := \varphi_t(U_t)$ which is independent of the occurrence date $t$ of the event. The discrete observation probabilities are easily extracted from this distribution using the relation
\begin{align}
	p_{t, s} &= P(U_t \in [s-t, s - t + 1)) \label{eq:formulaP} \\
			 &= F_{\tilde{U}}\left(\sum_{i=1}^{s-t+1} \alpha_{t, t+i-1}\right) - F_{\tilde{U}}\left(\sum_{i=1}^{s-t} \alpha_{t, t+i-1}\right) \nonumber.
\end{align}

Under the time change transformation the constraints \eqref{eq:constraint} become
\begin{equation*}
	\alpha_{t, s} \geq 0, \quad \forall t,s \quad \text{and} \quad \sum_{s \geq t} \alpha_{t, s} =  \infty, \quad \forall t.
\end{equation*}
We specify a regression model for the daily observation exposure as a function of covariates. We set
\begin{equation*}
	\log(\alpha_{t, s}) =  \vectorF{x}_{t, s}^{'} \cdot \vectorF{\gamma},
\end{equation*}
for a vector $\vectorF{x}_{t, s}$ of covariates related to observing on date $s$ an event that occurred on date $t$ and the corresponding parameter vector $\vectorF{\gamma}$. In contrast with classical regression methods, the reporting probabilities $p_{t, s}$ not only depend on the characteristics of the observation date, but instead take the full history between the event occurrence and observation date into account through the time change strategy.

Figure~\ref{figure:timeChangeFull} illustrates this time change. Since less claims get reported during the weekend, we model observation exposure as a function of the reporting day of the week. The time change then assigns lower observation exposures to Saturday and Sunday, hereby transforming the continuous distribution from Figure~\ref{figure:timeChangeContinuous} into a time-changed distribution that can be modeled using standard loss distributions.

\subsection{Calibration} \label{section:calibration}
Our approach divides the observation delay model into two components. The time change transformation $\varphi_t$ defined in $\eqref{eq:transformation}$ captures the heterogeneity in the observation process. This transformation is expressed by the daily observation exposures, which require the calibration of the regression parameters $\vectorF{\gamma}$. The time transformed observation delay $\tilde{U}$ is modeled with a simple parametric probability distribution, where the data will assist us in choosing the best candidate. We optimize the loglikelihood in $\eqref{eq:likelihoodgeneric}$ with respect to $\vectorF{\gamma}$, i.e. we maximize
\begin{align*}
	\ell(\vectorF{\gamma} ; \vectorF{\chi} ) &= \sum_{t = 1}^\tau \sum_{s = t}^\tau N_{t, s} \cdot \log\left[F_{\tilde{U}}\left(\sum_{v=t}^{s} \alpha_{t, v} \right) - F_{\tilde{U}} \left(\sum_{v=t}^{s-1} \alpha_{t, v} \right) \right] \\
 &\phantom{=} - \sum_{t=1}^\tau N_t^{\Rep}(\tau) \cdot \log\left[F_{\tilde{U}} \left(\sum_{v=t}^\tau \alpha_{t, v} \right)\right],
\end{align*}
with $\alpha_{t, v} = \exp \left( \vectorF{x}_{t, v}^{'} \cdot \vectorF{\gamma}\right)$. Online appendix~\ref{appendix:fittingProcedure} describes an optimization strategy for this loglikelihood that is applicable to any sufficiently smooth distribution $F_{\tilde{U}}(\, \cdot \,)$. The described strategy is generic and does not immediately take properties from the chosen distribution into account. Significant reductions in computation time can be obtained when $\tilde{U}$ follows a standard exponential distribution. The loglikelihood then becomes
\begin{align} \label{eq:loglikelihoodExp}
	\ell(\vectorF{\gamma}; \vectorF{\chi} ) &= -\sum_{t=1}^\tau \sum_{s=t}^{\tau} N_{t, s} \cdot \left( \sum_{v=t}^{s-1} \alpha_{t, v} - \log \left( 1 - \exp\left( -\alpha_{t, s} \right) \right) \right) \\
											&\phantom{=} - \sum_{t=1}^{\tau} N_t^{\Rep}(\tau) \cdot \log\left( 1 - \exp \left( - \sum_{v=t}^\tau \alpha_{t, v} \right)\right). \nonumber
\end{align}
The first line in $\eqref{eq:loglikelihoodExp}$ is a sum in which each term depends on a single observation exposure, $\alpha_{t, s}$. Since this facilitates computing first and second order derivatives with respect to the reporting exposure, this results in a lower computation time.

\subsection{Predicting the number of hidden events} \label{section:estimation}
At the evaluation date $\tau$ we predict the number of events from past occurrence dates $t$ that will be observed on future dates $s$. Hence our focus is on
\begin{equation*}
	N_{t, s}, \quad \text{for } t \leq \tau \text{ and } s > \tau.
\end{equation*}
We aggregate these future daily observation counts to find the total number of hidden events
\begin{equation*}
		N^{\Hidden}(\tau) = \sum_{t=1}^\tau N_t^{\Hidden}(\tau) = \sum_{t=1}^\tau \sum_{s=\tau + 1}^\infty N_{t,s}.
\end{equation*}
Following the Poisson assumption in $\eqref{eq:poissonThinning}$ each random variable $N_{t, s}$ is independently Poisson distributed with mean
\begin{equation*}
	E(N_{t, s}) = \lambda_t \cdot p_{t,s}.
\end{equation*}
The observation delay model developed in Section \ref{section:dailyReportingWeights} provides estimates for the observation probabilities $p_{t, s}$, see $\eqref{eq:formulaP}$
\begin{equation*}
	\hat{p}_{t, s} = P(\tilde{U} \in [\varphi_t(s-t), \varphi_t(s-t+1)) \mid \hat{\vectorF{\gamma}}).
\end{equation*}
In $\eqref{eq:zeta}$ we proposed a pragmatic, non-parametric estimator for the claim occurrence intensity on date $t$, namely
\begin{equation}
	\hat{\lambda}_t = \frac{N_t^{\Observed}(\tau)}{\hat{p}^{\Observed}_t(\tau)}. \label{eq:estNt}
\end{equation}
This estimator depends only on the observed events and the estimated observation delay distribution. This is an advantage when the event generating process is volatile. For dates with unexpectedly many events the number of observations will be higher and thus we correctly predict more event occurrences. On the downside, $\eqref{eq:estNt}$ is less reliable for recent dates when the denominator is close to zero or when the number of daily events is low. When the data set is small, the non-parametric estimator can be replaced by a parametric estimator following the strategy outlined in \cite{Bonetti2016} and \cite{Verbelen2017}. In a parametric framework the estimator for the occurrence intensity may include the daily risk exposure, expressed as the number of policies in effect on a day. Including risk exposure increases the robustness of parametric models to evolutions in the portfolio size and may potentially improve the predictive performance of the model.

\section{Case-study: reporting delay dynamics in insurance} \label{section:caseStudy}

\subsection{Data characteristics} \label{section:dataDescription}

We illustrate our approach with the analysis of a liability insurance data set from the Netherlands. The same data is studied in \cite{PigeonAntonioDenuit2013}, \cite{Pigeon2014} and \cite{Godecharle2015} with focus on calculating reserves in discrete time, \cite{AntonioPlat2014} model reserves in continuous time and \cite{Verbelen2017} who propose a model for the number of hidden claim counts at a daily level. The data registers \num{506235} claims related to insured events that occurred and were reported between July, 1996 and August, 2009. From these claims, we remove 75 observations with a reporting date prior to the accident date and 559 claims that are the result of transitions in the reporting system. We focus on the occurrence date of accidents and the corresponding reporting delay in days, i.e.~the time (in days) between occurrence of the accident and reporting or filing of the claim to the insurer. To avoid losing valuable insights by aggregation, we study the data at a daily level. This is the most granular timescale at which the data is available.

\begin{figure}[ht!]
\hfill
\includegraphics[width=.95\textwidth]{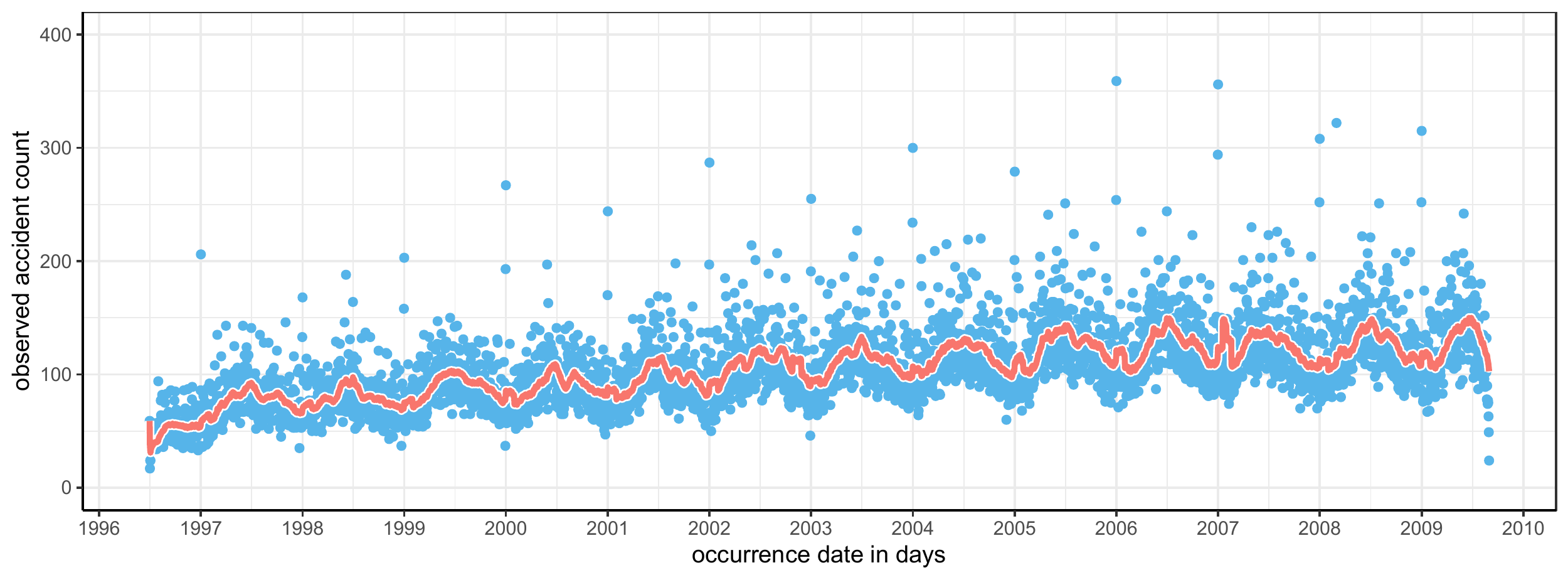}
\hfill
\caption{Daily number of accidents that occurred between July, 1996 and August, 2009 and were reported before August, 2009. The solid line shows the moving average of occurred accidents, calculated over the latest 30 dates. Two outliers are not shown on the graph: October 27, 2002 (456 accidents) and January 18, 2007 (818 accidents).}
\label{figure:ClaimCount}
\end{figure}

\paragraph{Occurred accidents}
Figure~\ref{figure:ClaimCount} shows the daily number of accidents that occurred between July, 1996 and August, 2009 and initiated a claim reported to the insurance company before August 31, 2009. Since only claims reported before August 31, 2009 are observed, we see a decrease in observed event counts for the most recent dates which have a substantial number of unreported claims. Two outliers are not shown in this plot, namely 456 accidents on October 27, 2002 and 818 accidents on January 18, 2007. Both outliers correspond to a storm in the Netherlands causing many insured events.\footnote{Details (in Dutch) about the storms by the royal national meteorological institute of the Netherlands (KNMI): \url{https://knmi.nl/over-het-knmi/nieuws/storm-van-27-oktober-2002-was-zwaarste-in-twaalf-jaar} and \url{https://knmi.nl/over-het-knmi/nieuws/de-zware-storm-kyrill-van-18-januari-2007} } The red line in this figure shows the moving average of the number of occurrences, calculated over the latest 30 days. This trend reveals a seasonal pattern in the occurrence process with more events occurring during the summer months. The trend slightly increases over time due to an increase in portfolio size.  Several of the outlying observations in Figure~\ref{figure:ClaimCount} correspond to occurrences on the first of January as indicated by the vertical gray bars at the beginning of each year.

\begin{figure}[ht!]
\hfill
\includegraphics[width=.95\textwidth]{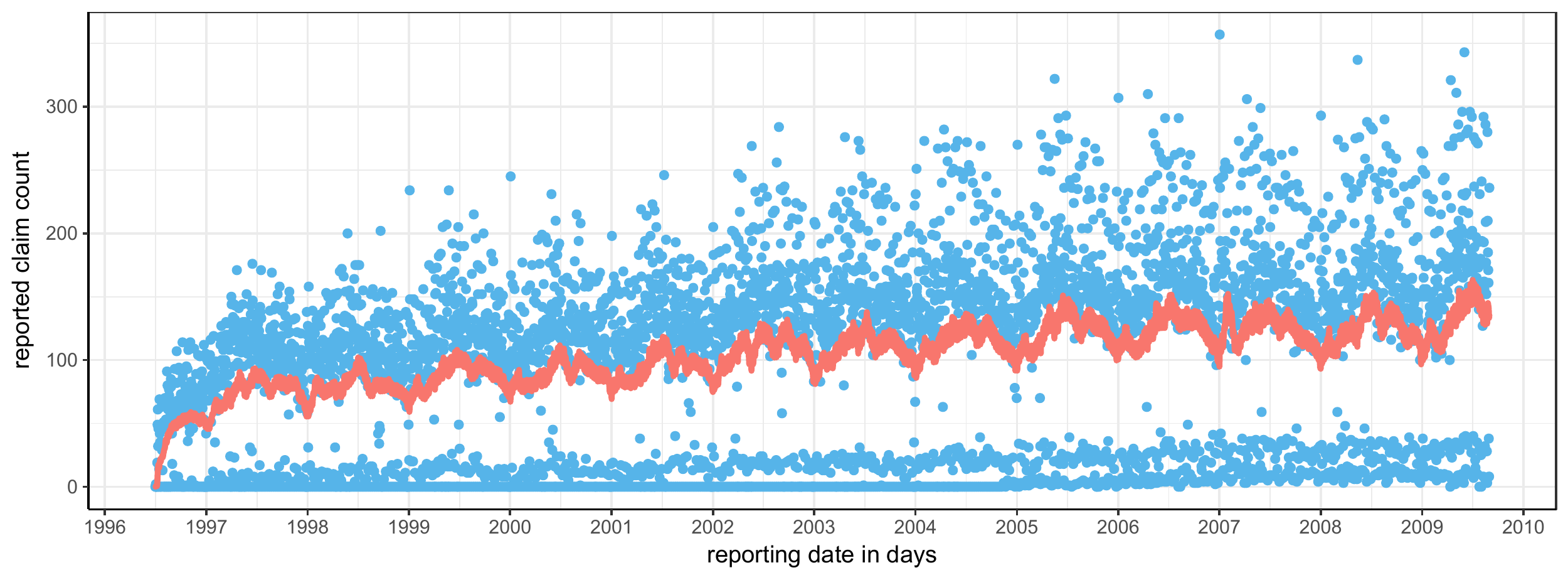}
\hfill
\caption{Daily number of claims that were reported on each date between July, 1996 and August, 2009. The solid line shows the moving average of reported claims, calculated over the latest 30 dates.}
\label{figure:ReportCount}
\end{figure}

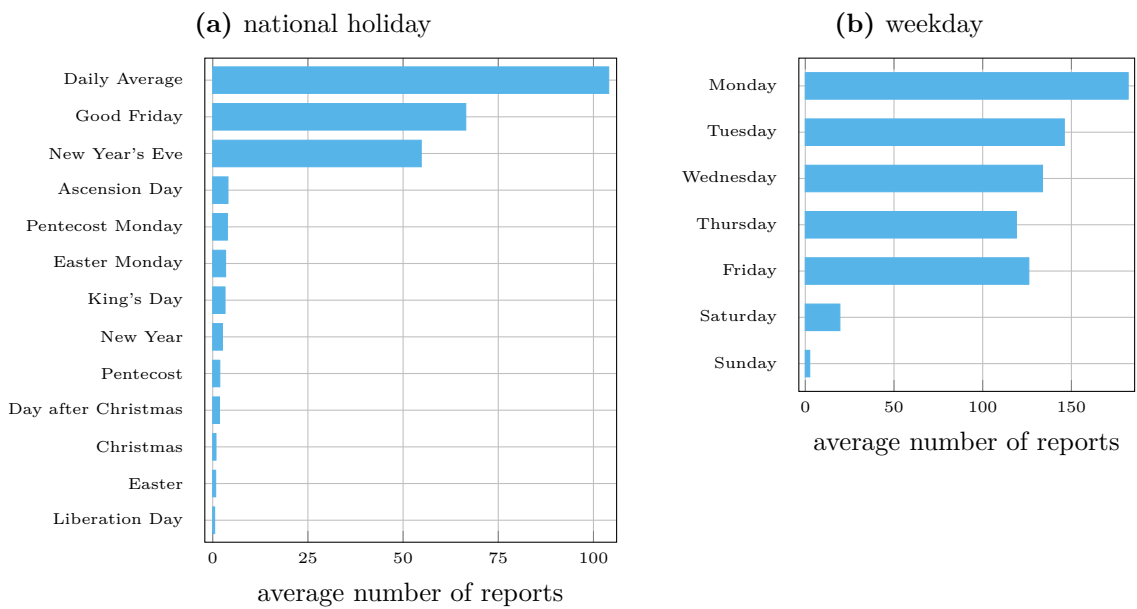
\begin{figure}[ht!]
\begin{subfigure}[t]{.520\textwidth}
\centering
\caption{national holiday}
\label{figure:CalendarEffectHoliday}
\begin{tikzpicture}
  \begin{axis}[
    xbar, xmin=0,
    width=7cm, height=8cm,
    xlabel={\small average number of reports},
    symbolic y coords={Liberation Day, Easter, Christmas, Day after Christmas, Pentecost, New Year, King's Day, Easter Monday, Pentecost Monday, Ascension Day, New Year's Eve, Good Friday, Daily Average},
    ytick=data,
    xtick={0, 25, 50, 75, 100},
    ymajorgrids,
    xmajorgrids,
    ticklabel style = {font=\tiny},
    ytick style={draw=none},
    enlarge y limits=0.05, enlarge x limits=0.02,
    ]
    \addplot[color=graphBlue, fill=graphBlue] coordinates {
		(0.50,Liberation Day)
		(0.77,Easter)
		(0.85,Christmas)
		(1.77,Day after Christmas)
		(1.85,Pentecost)
		(2.54,New Year)
		(3.23,King's Day)
		(3.38,Easter Monday)
		(3.85,Pentecost Monday)
		(4,Ascension Day)
		(54.76,New Year's Eve)
		(66.46,Good Friday)
		(104,Daily Average)};
  \end{axis}
\end{tikzpicture}
\end{subfigure}%
\begin{subfigure}[t]{.460\textwidth}
\centering
\caption{weekday}
\label{figure:CalendarEffectWeekday}
\begin{tikzpicture}
  \begin{axis}[
    xbar, xmin=0,
    width=6cm, height=6cm,
    xlabel={\small average number of reports},
    symbolic y coords={Sunday, Saturday, Friday, Thursday, Wednesday, Tuesday, Monday},
    ytick=data,
    xtick={0, 50, 100, 150},
    ymajorgrids,
    xmajorgrids,
    ticklabel style = {font=\tiny},
    ytick style={draw=none},
    enlarge y limits=0.10, enlarge x limits=0.02,
    ]
    \addplot[color=graphBlue, fill=graphBlue] coordinates {
		(182,Monday)
		(146,Tuesday)
		(133.7,Wednesday)
		(119,Thursday)
		(126,Friday)
		(19.5,Saturday)
		(2.5,Sunday)};
  \end{axis}
\end{tikzpicture}
\end{subfigure}

\caption{Average number of reported claims on {\bf (a)} national holidays and {\bf (b)} weekdays, calculated over all claims that occurred and were reported between July, 1996 and August, 2009.}
\label{figure:CalendarEffect}
\end{figure}

\paragraph{\revision{Reported claims}}
Figure~\ref{figure:ReportCount} shows the daily number of claims reported between July 1996 and August 2009. Again the red line shows the moving average of the number of reported claims, calculated over the latest 30 days. The seasonality in event counts observed in Figure~\ref{figure:ClaimCount} leads to a similar seasonal pattern in reported claim counts, though with a slight lag due to the delay in reporting a claim. Figure~\ref{figure:ReportCount} reveals two regimes of reporting. On most dates many claims get reported, but there is a substantial number of dates on which few or almost no claims are reported. These dates with few reports correspond to the weekend (Saturday, Sunday) and national holidays.\footnote{List of national holidays in the Netherlands: \url{http://www.officeholidays.com/countries/netherlands/} } This separation in two regimes is not the case for the occurrence process, since accidents continue to occur during the weekend and on holidays. We further illustrate these calendar day effects, where reporting is substantially reduced on specific dates, in Figure~\ref{figure:CalendarEffect}. The left hand side lists the average number of reported claims between July, 1996 and August, 2009 on ten national holidays during which all businesses are closed. These averages are compared with the overall daily average of reported claim counts over the observation period. This shows that reporting is strongly reduced on national holidays. We include two non-official holidays, New Year's Eve and Good Friday. These dates show a slight reduction in reporting because many people take a day off from work. The reporting behavior on weekdays is shown in Figure~\ref{figure:CalendarEffectWeekday}. During the weekend and especially on Sunday the number of reports is reduced. These calendar day effects motivate a model for IBNR claim counts at a daily level, capable of incorporating the weekday and holiday effect observed in our empirical analysis.

\begin{figure}
\hfill
\hfill
\begin{subfigure}[t]{.32\textwidth}
\caption{All claims}\label{figure:delayDistributionThreeWeeksA}
\includegraphics[width=\textwidth]{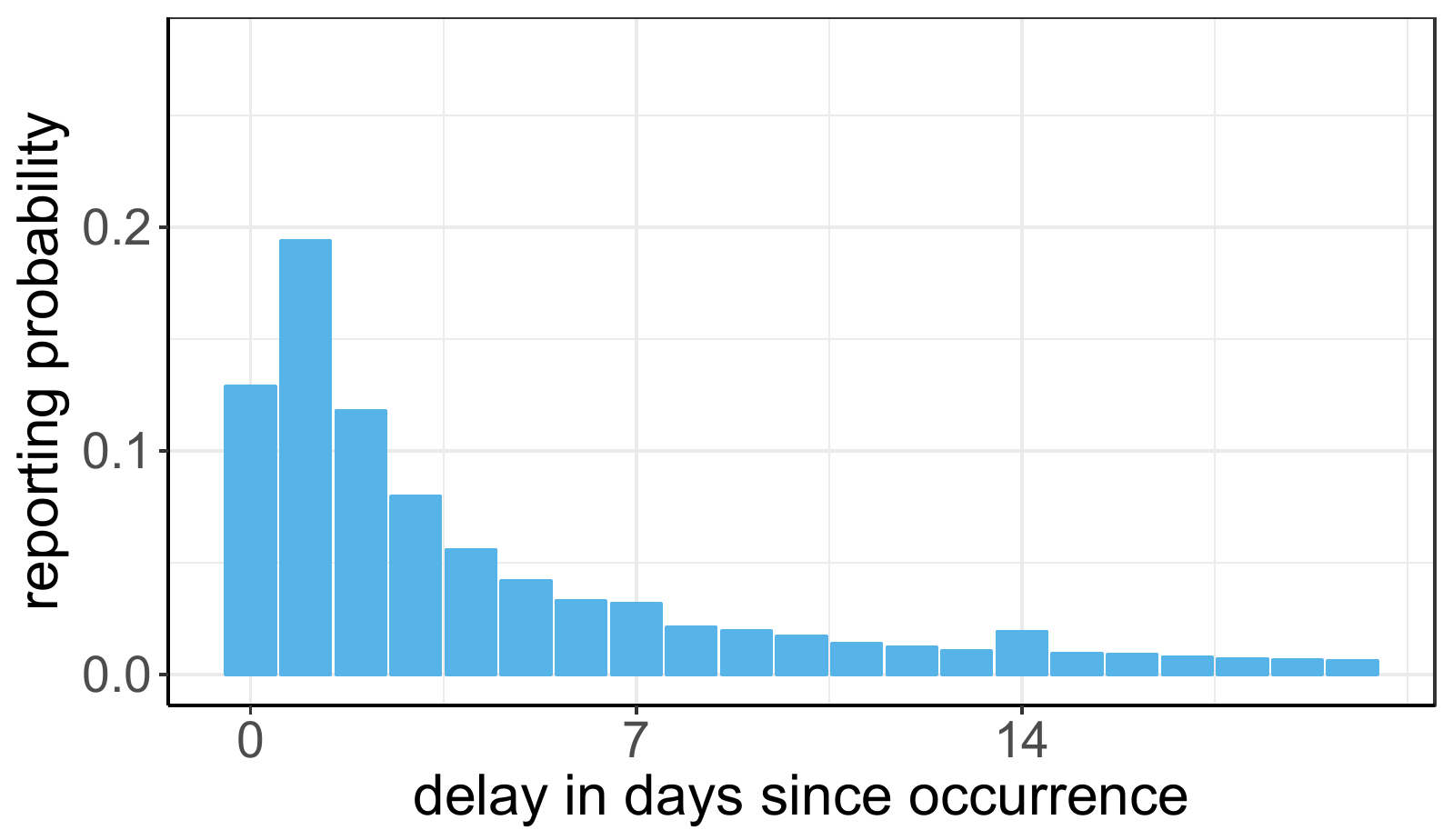}
\end{subfigure}
\hfill
\begin{subfigure}[t]{.32\textwidth}
\caption{Monday}\label{figure:delayDistributionThreeWeeksB}
\includegraphics[width=\textwidth]{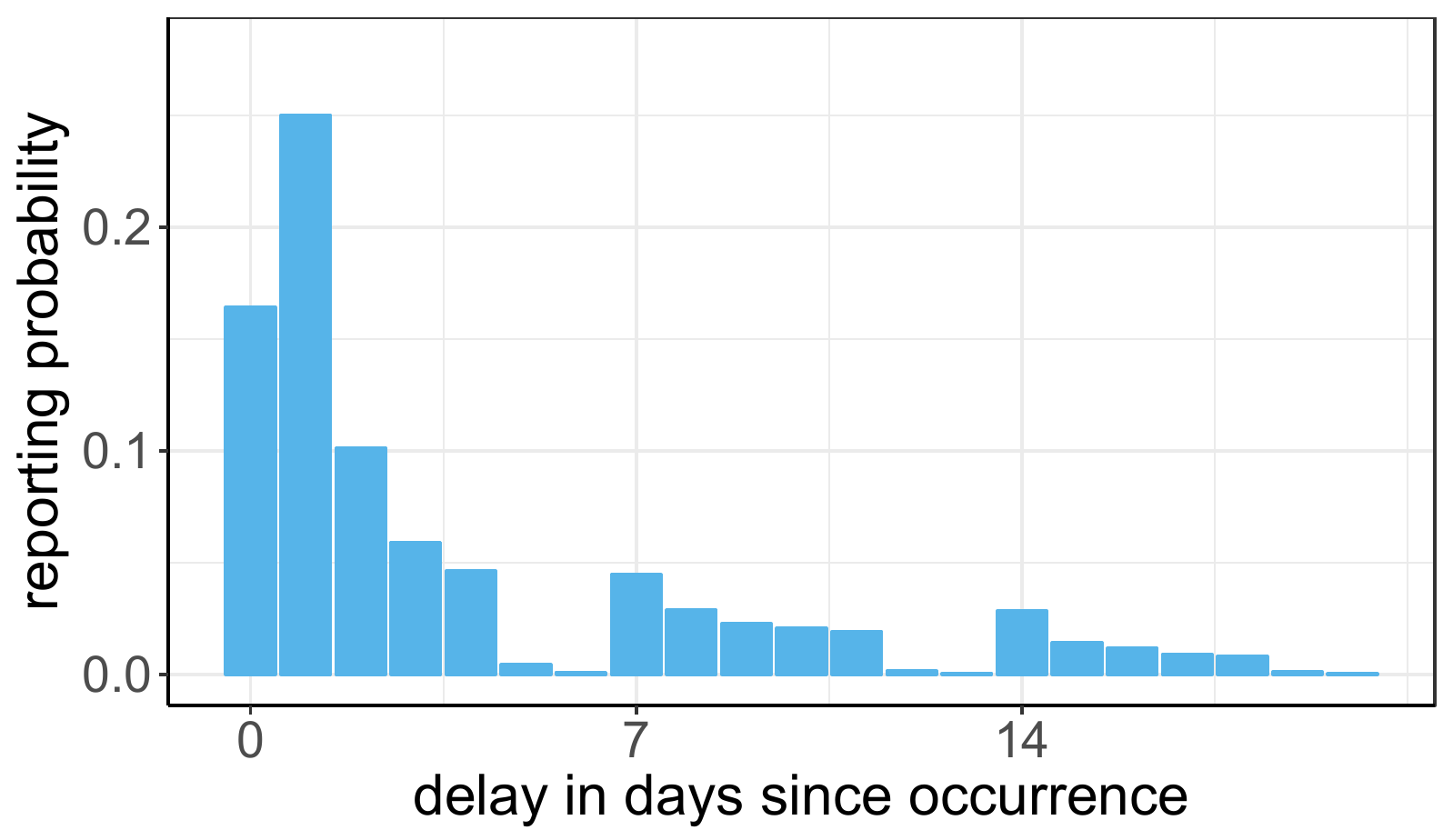}
\end{subfigure}
\hfill
\begin{subfigure}[t]{.32\textwidth}
\caption{Thursday}\label{figure:delayDistributionThreeWeeksC}
\includegraphics[width=\textwidth]{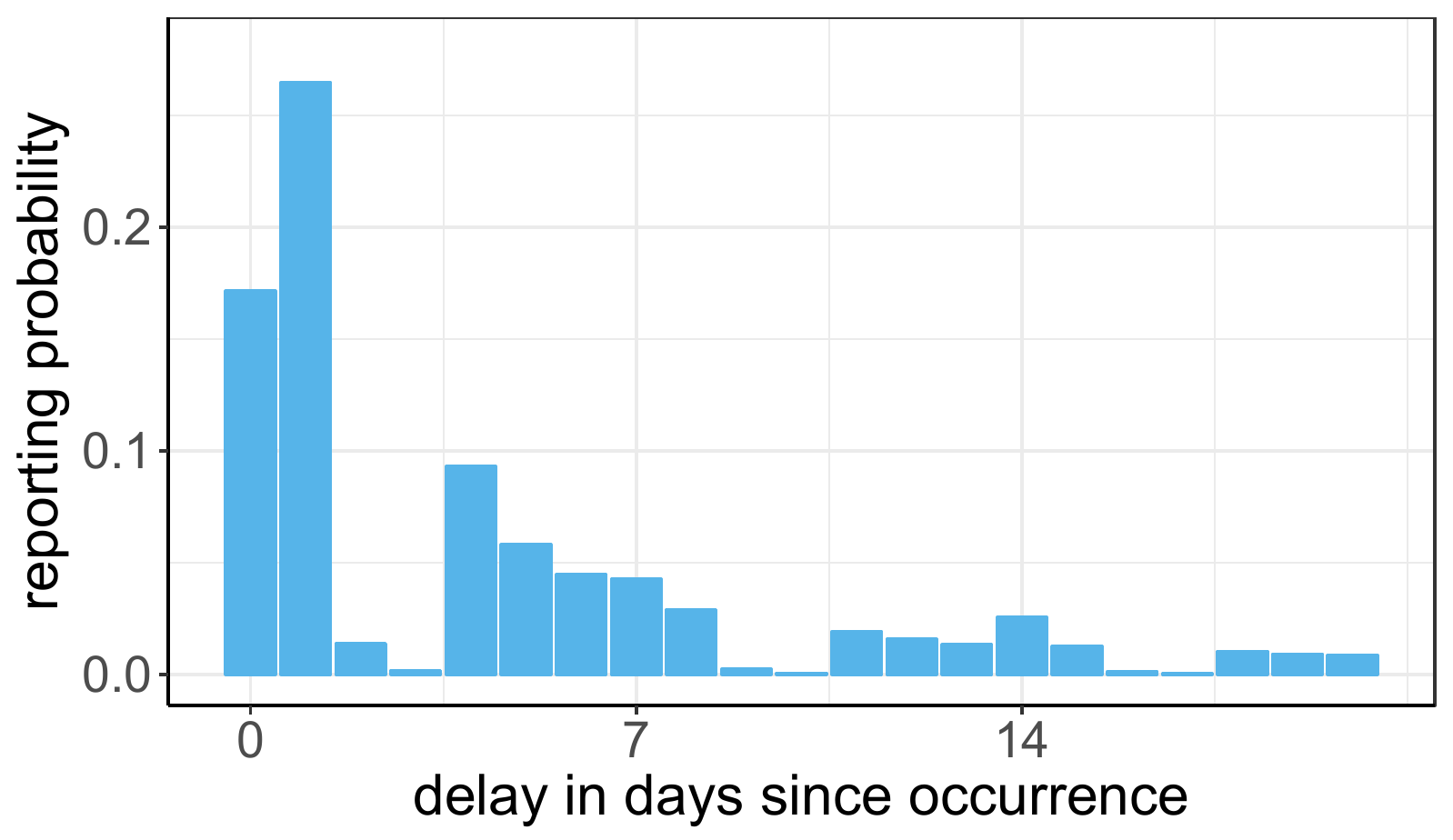}
\end{subfigure}
\hfill
\caption{Empirical reporting delay distribution in days over the first three weeks after the occurrence of the claim using {\bf (a)} all claims, {\bf (b)} claims that occurred on a Monday and {\bf (c)} claims that occurred on a Thursday.}
\label{figure:delayDistributionThreeWeeks}
\end{figure}

\paragraph{Reporting delay}
Figure~\ref{figure:delayDistributionThreeWeeks} illustrates the empirical reporting delay distribution in days over the first three weeks after the occurrence of the insured event. The empirical probability of reporting peaks the day after the claim occurred and strongly decreases afterwards. The increase in reporting after exactly fourteen days is most likely a consequence of data quality issues, where insureds who no longer recall the exact occurrence date report that the accident happened two weeks ago. The same effect to a lesser degree is visible after exactly one week. 
Figure~\ref{figure:delayDistributionThreeWeeksB} and Figure~\ref{figure:delayDistributionThreeWeeksC} show the empirical reporting delay distribution constructed using only accidents that occurred on Monday and Thursday, respectively. This reveals the effect of the occurrence's day of the week on the reporting delay distribution. An accident that happened on a Monday has a decreased probability of reporting after six or seven days, since these delays correspond to Saturday and Sunday, respectively. Accidents that occurred on a Thursday show the same pattern of reporting delay, but the weekend then corresponds to a different delay. The effect of the weekend is no longer visible in the empirical distribution using all claims (Figure~\ref{figure:delayDistributionThreeWeeksA}), since the weekend then no longer corresponds to a specific reporting delay.

\paragraph{The number of hidden events}

\begin{figure}
\captionsetup[subfigure]{labelformat=empty}
\hfill
\includegraphics[width=0.95\textwidth]{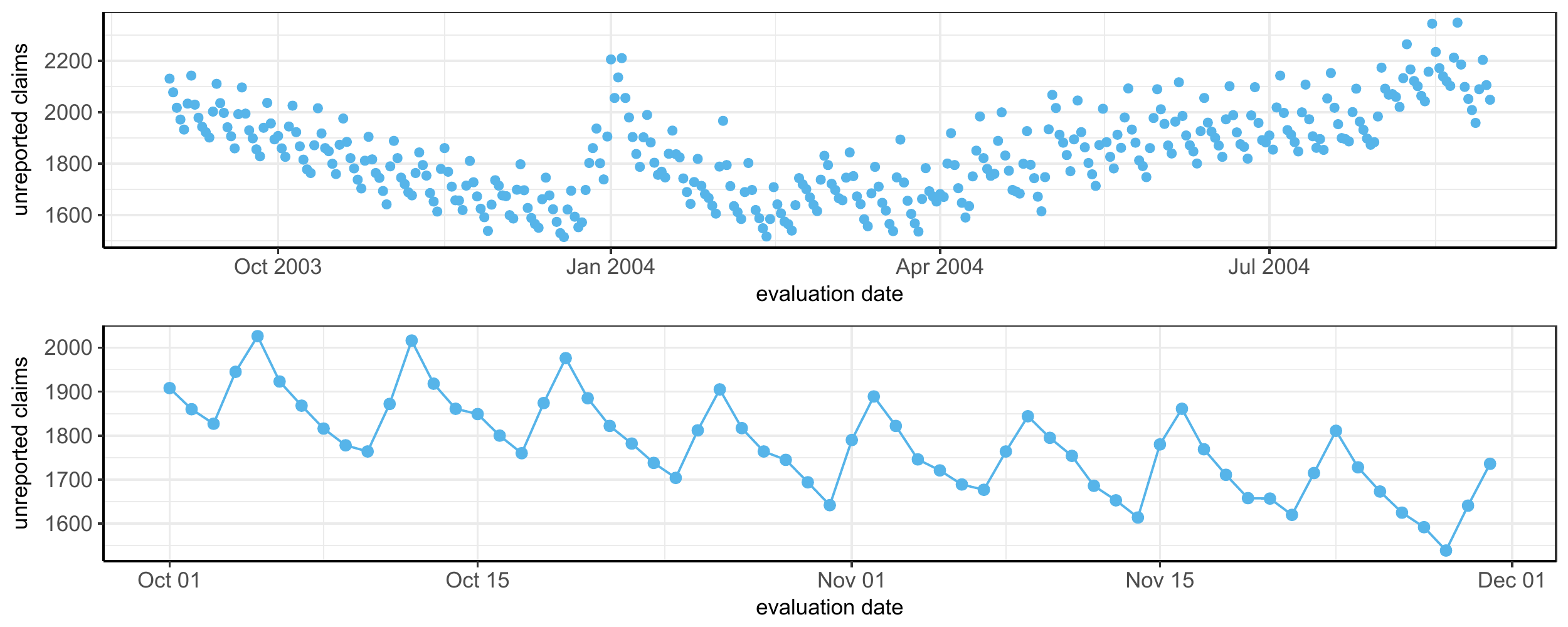}
\hfill
\caption{Number of unreported claims at each evaluation date between September 2003 and August 2004. These are the number of claims that occurred before this date, but were reported afterwards (but before the end of the observation period, i.e. August 31, 2009). The bottom panel zooms in on evaluation dates in October and November, 2003.}
\label{figure:totalIBNR}
\end{figure}

The evaluation date refers to the date on which the insurer computes the reserve. In practice this date is often the last day of a quarter or the financial year. Figure~\ref{figure:totalIBNR} uses a rolling evaluation date to illustrate the daily number of IBNR claims. For each evaluation date we show the number of claims corresponding to insured events that occurred before this date but were reported afterwards (and before August 31, 2009, the last day of our observation period). The top panel of Figure~\ref{figure:totalIBNR} shows the daily number of IBNR claims on each evaluation date between September 1, 2003 and August 31, 2004. The number of unreported claims varies throughout the year with more unreported claims in the summer, when more accidents occur. IBNR counts peak around the start of the new year since many accidents occur on the first of January and reporting is slow due to a clustering of holidays. The bottom panel of Figure~\ref{figure:totalIBNR} zooms in on the unreported claims between October 1, 2003 and November 30, 2003. Large fluctuations in unreported claims appear when we evaluate IBNR on a daily basis. These movements follow a seven day pattern where five days of decrease in IBNR are followed by two days of strong upward movement. These upward moves correspond to the weekend when many new insured events occur, but almost no events get reported.

\subsection{Model specification} \label{section:modelliability}

We opt for computational efficiency and model the time-changed reporting delay $\tilde{U}$ with an exponential distribution. The reporting exposures include six effects and are structured as
\begin{align}
\alpha_{t, s} &=
	\alpha^{\texttt{occ.\,dom}}_t
\cdot
	\alpha^{\texttt{occ.\,month}}_t
\cdot
	\alpha^{\texttt{rep.\,holiday}}_{s}
\cdot
	\alpha^{\texttt{rep.\,month}}_{s}
\cdot
	\alpha^{\texttt{rep.\,dow,\,first week}}_{s, s-t}
\cdot
	\alpha_{s-t}^{\texttt{delay}} \label{eq:modelParametersReporting} \\
					&= \exp\left(
	 \covariateDisplay{(\vectorF{x}_{t}^{\texttt{occ.\,dom}})^{'} \cdot \gamma^{\texttt{occ.\,dom}}}{\vectorF{\gamma}^{\texttt{occ.\,dom}} \cdot \texttt{dom(t)}}
+
	\covariateDisplay{(\vectorF{x}_{t}^{\texttt{occ.\,month}})^{'} \cdot \gamma^{\texttt{occ.\,month}}}{\vectorF{\gamma}^{\texttt{occurrence month}} \cdot \texttt{month(t)}}
\nonumber \right.\\ & {\color{white}\, = \exp (} +
	\covariateDisplay{(\vectorF{x}_{s}^{\texttt{rep.\,holiday}})^{'} \cdot \gamma^{\texttt{rep.\,holiday}}}{\vectorF{\gamma}^{\texttt{holiday}} \cdot \texttt{holiday(s)}} +
	\covariateDisplay{(\vectorF{x}_{s}^{\texttt{rep.\,month}})^{'} \cdot \gamma^{\texttt{rep.\,month}}}{\vectorF{\gamma}^{\texttt{reporting month}} \cdot \texttt{month(s)}}
\nonumber \\ &{\color{white}\, = \exp (} +
	\covariateDisplay{(\vectorF{x}_{s, s-t}^{\texttt{rep.\,dow,\,first week}})^{'} \cdot \gamma^{\texttt{rep.\,dow,\,first week}}}{\vectorF{\gamma}^{\texttt{dow, first week}} \cdot \texttt{dow(s) * first week(s-t)}}
+
	\left. \covariateDisplay{(\vectorF{x}_{s-t}^{\texttt{delay}})^{'} \cdot \gamma^{\texttt{delay}}}{\vectorF{\gamma}^{\texttt{delay}} \cdot \texttt{delay(s-t)}}\right) . \nonumber
\end{align}
We model the impact of the occurrence date on the reporting delay by incorporating effects for the day of the month $\alpha^{\texttt{occ.\,dom}}_t$ and the month $\alpha^{\texttt{occ.\,month}}_t$ on which the accident occurs. The holiday effect in Figure~\ref{figure:CalendarEffectHoliday} is modeled by $\alpha_{s}^{\texttt{rep.\,holiday}}$, which distinguishes between national and unofficial holidays. Seasonal variations in reporting are captured by $\alpha^{\texttt{rep.\,month}}_{s}$, which scales reporting exposure based on the month in which the claim is reported. An interaction effect $\alpha_{s, s-t}^{\texttt{rep.\,dow,\,first week}}$ estimates the reporting exposure for combinations of a reporting delay in the first week ($s-t = 0, 1, \ldots, 6$) and the day of the week on which the claim is reported. Separate weekday parameters are estimated for delays of more than one week, $s-t \geq 7$. As such, we capture the weekday effect from Figure~\ref{figure:CalendarEffectHoliday} with additional flexibility in the first week after the claim occurs. Finally, $\alpha_{s-t}^{\texttt{delay}}$ partitions the time elapsed since the accident occurred in 23 bins according to the strategy specified in online Appendix~\ref{section:AdaptingExponential}. These bins adapt the tail of the distribution as well as increase the probability of reporting after 14, 30 and 365 days.

\subsection{Results} 

\subsubsection{Parameter estimates} \label{section:modelParameters}

\begin{figure}[ht!]
\captionsetup[subfigure]{labelformat=empty}
\centering

\begin{subfigure}{.95\textwidth}
\includegraphics[width=\textwidth]{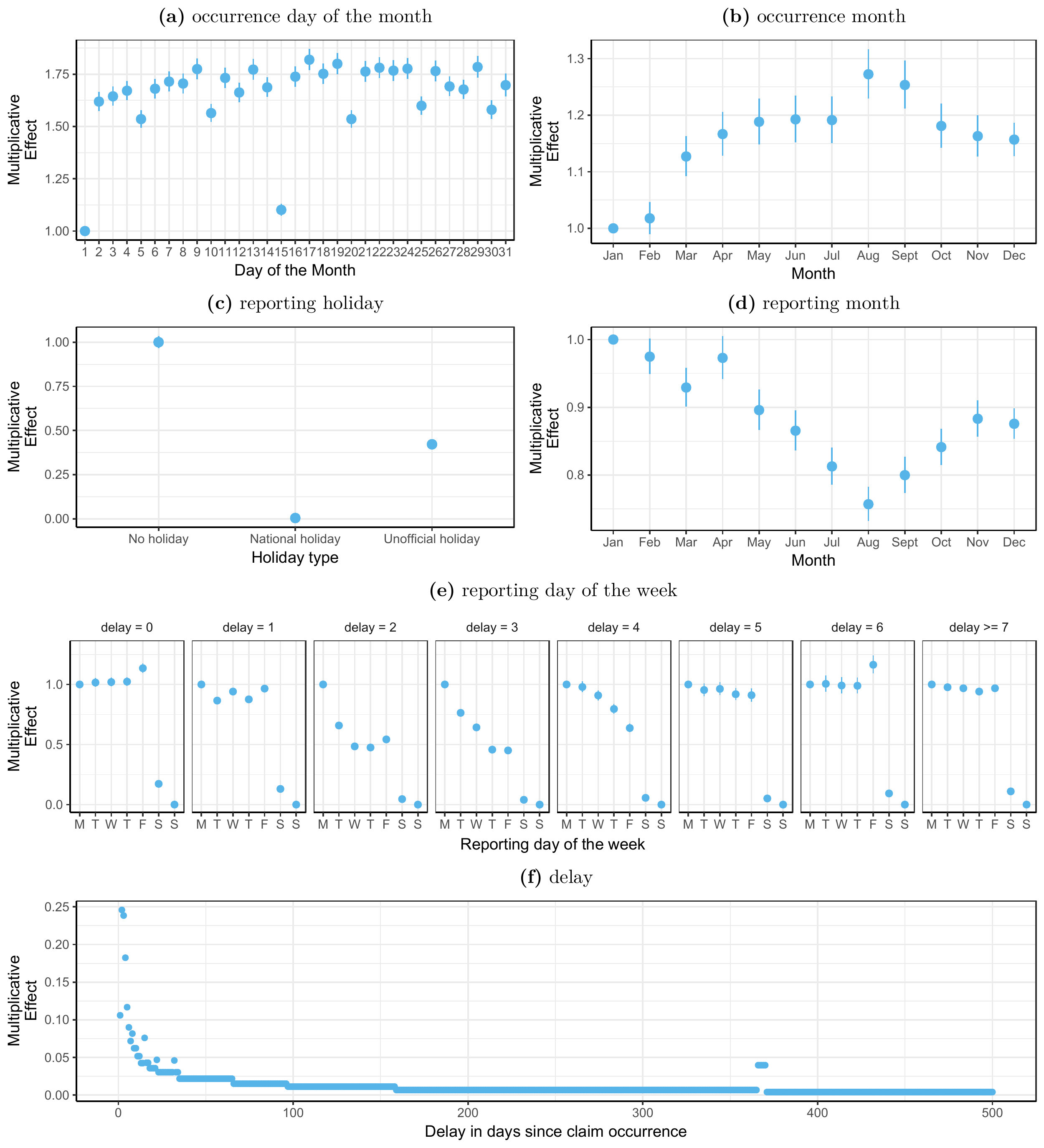}
\phantomcaption \label{fig:paramdom}
\phantomcaption \label{fig:paramoccmonth}
\phantomcaption \label{fig:paramholiday}
\phantomcaption \label{fig:paramrepmonth}
\phantomcaption \label{fig:paramdow}
\phantomcaption \label{fig:paramdelay}
\end{subfigure}

\caption{Maximum likelihood estimates with $95\%$-confidence intervals for the reporting exposure parameters $\exp(\vectorF{\gamma})$ in \eqref{eq:modelParametersReporting}.}
\label{figure:maximumLikelihoodEstimatesReporting}
\end{figure}

We estimate the model parameters by maximizing the loglikelihood in $\eqref{eq:loglikelihoodExp}$ using 8 years of data i.e.~all accidents that occurred and were reported between July 1, 1996 and September 5, 2004. The resulting training data set contains \num{274187} reported claims, for which we model the reporting process using \num{125} parameters. Figure~\ref{figure:maximumLikelihoodEstimatesReporting} shows the maximum likelihood estimates for the reporting exposure parameters $\exp(\vectorF{\gamma})$ in \eqref{eq:modelParametersReporting}. Together with these point estimates we plot $95\%$-confidence intervals derived from the Fisher information matrix for $\vectorF{\gamma}$.

\paragraph{Occurrence day of month} Figure~\ref{fig:paramdom} shows the effect of the day of the month on which the accident occurred. Reporting exposure is lower for accidents that occur on the first or fifteenth of the month, which implies that accidents from these days have a longer reporting delay. This is most likely the result of data quality issues. Insureds who report a claim with a long reporting delay might no longer remember the exact occurrence date of the corresponding accident, which leads them to register the occurrence date at the start (first) or middle (fifteenth) of the month. This creates an increase in the average reporting delay for events that occurred on the first and fifteenth of the month. The same effect to a lesser degree is visible on the 5th, 10th, 20th, 25th and 30th of the month.

\paragraph{Month} Two month effects are included in the reporting exposure structure. Figure~\ref{fig:paramoccmonth} shows the effect for $\exp(\vectorF{\gamma}^{\texttt{occ.\,month}})$ which considers the month in which the accident occurs. These parameters indicate that reporting is slower for accidents that occurred around the beginning of the year (January, February) and faster in the summer. Figure~\ref{fig:paramrepmonth} visualizes the parameters for the reporting month, $\exp(\vectorF{\gamma}^{\texttt{rep.\,month}})$. We observe a reduction in reporting exposure during the summer months. Slightly counterintuitive, we find that the parameters $\gamma^{\texttt{occ.\,month}}$ and $\gamma^{\texttt{rep.\,month}}$ largely offset each other for accidents that occur and get reported in the same calendar month. When combining these effects, the reduction in reporting exposure during the summer is mostly noticeable for claims that occurred before the summer months.

\paragraph{Holiday} Figure~\ref{fig:paramholiday} shows the effect of holidays on reporting \weight{}. Hardly any claim gets reported on national holidays and the reporting probability is reduced by more than $50\%$ on unofficial holidays (Good Friday and New Year's Eve). These estimates are of the same magnitude as the effects found in the empirical analysis in Figure~\ref{figure:CalendarEffect}.

\paragraph{Reporting day of the week} We include the day of the week effect in the reporting \weight{} specification \eqref{eq:modelParametersReporting} through an interaction between the time elapsed after the accident occurred $s-t$ and the day of the week on which the claim is reported. Figure~\ref{fig:paramdow} shows a grouping of the estimated coefficients based on the time elapsed since the occurrence of the accident. For all delays we notice a reduction in reporting exposure during the weekend, with few reports on Saturday and almost no reports on Sunday. This interaction is important as the estimated parameters differ strongly based on the delay considered. For example, accidents that occur on Friday or Saturday are often reported on the next Monday, which corresponds to a delay of two and three days respectively. Since Monday is the reference level, the fitted parameters for other weekdays are lower at these delays. The right most panel in Figure~\ref{fig:paramdow} shows the effect of the reporting day of the week for delays beyond one week. For these longer delays, all working days (\texttt{Mon} - \texttt{Fri}) have a similar reporting exposure.

\paragraph{Delay} Figure~\ref{fig:paramdelay} shows the evolution of the reporting \weight{} component $\exp(\vectorF{\gamma}^{\texttt{delay}})$ in \eqref{eq:modelParametersReporting} as a function of the time elapsed since the accident occurred. This effect scales the reporting probability at specific delays such that the time-changed reporting delay $\tilde{U}$ better resembles an exponential distribution. We identified 23 bins upfront based on the strategy of online appendix~\ref{section:AdaptingExponential}. The first eight days after occurrence end up in separate bins. These short delays are important, since many claims get reported soon after their occurrence date. Moreover, Figure~\ref{fig:paramdelay} shows that the calibrated effect changes strongly for these delays. The model also contains bins to capture the increase in reporting probability for delays of exactly $14$, $21$ and $31$ days as well as for reporting after one year. The bin size widens when reporting delay increases. The final two bins $[158, 364]$ and $[370, \infty)$ let the model capture the tail of the distribution.

\subsection{Out-of-time predictions} \label{section:modelPredictions}
We predict the number of hidden events, i.e.~the IBNR claim count, following the strategy outlined in Section~\ref{section:estimation}. Because the non-parametric occurrence estimators are unreliable for recent event dates for which few events are observed, we propose a pragmatic approach to get around this drawbacks. Insurance companies use very specific evaluation dates when calculating reserves, such as the end of a quarter, semester or financial year. Typically the calculations are not performed at those exact evaluation dates, but a couple of days later (at the so-called computation date). Accordingly we predict the number of hidden events on August 31, 2004 using data until September 5, 2004. As such, the granular model predicts 2012.7 unreported claims on August 31, 2004, whereas the true number of IBNR claims (based on data until August 31, 2009) was 2049.

\begin{figure}
\hfill
\begin{subfigure}{.99\textwidth}
\includegraphics[width = \textwidth]{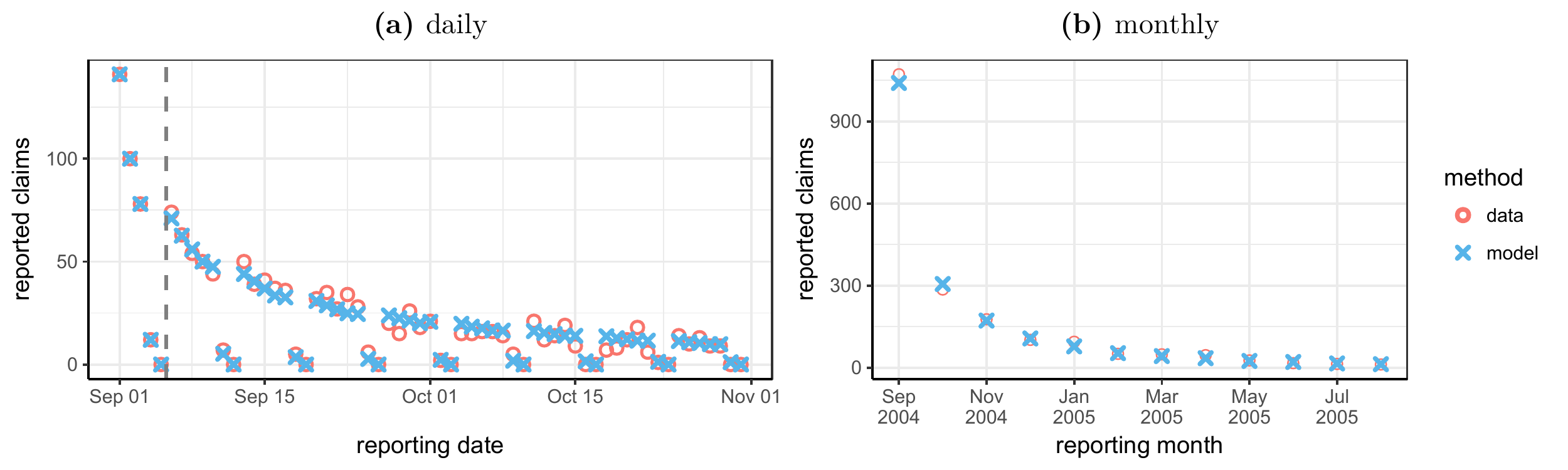}
\phantomcaption \label{fig:claimCountPredDaily}
\phantomcaption \label{fig:claimCountPredMonthly}
\end{subfigure}
\caption{Out-of-time prediction of the number of reported claims for accidents that occurred before August 31, 2004. These predictions are compared with the actual number of reported claims. {\bf (a)} Estimated at a daily level for the next two months. The dashed line indicates the last observed date (September 5, 2014). {\bf (b)} Estimates aggregated by reporting month for the next twelve months.}
\label{figure:claimCountPred}

\end{figure}

\paragraph{Future observation of hidden events} Our daily model splits the total IBNR point estimate of 2012.7 claims by future reporting date. Figure~\ref{fig:claimCountPredDaily} shows the estimated number of daily reported claims in September and October, 2004 for accidents that occurred before August 31, 2004. The dashed line in Figure~\ref{fig:claimCountPredDaily} indicates the computation date. We do not make predictions for dates falling before the computation date as this data is observed. The model accurately predicts the low report counts during the weekend. This is the merit of adding the day of the week effect in the reporting \weight{} model. Also the overall reporting pattern closely matches the observed values. Figure~\ref{fig:claimCountPredMonthly} aggregates these daily report counts by month. This figure shows the estimated number of reported claims in the first twelve months following August, 2004. In these months the observed and predicted IBNR counts are very similar.

\begin{figure}
\centering

\captionsetup[subfigure]{labelformat=empty}
\includegraphics[width = \textwidth]{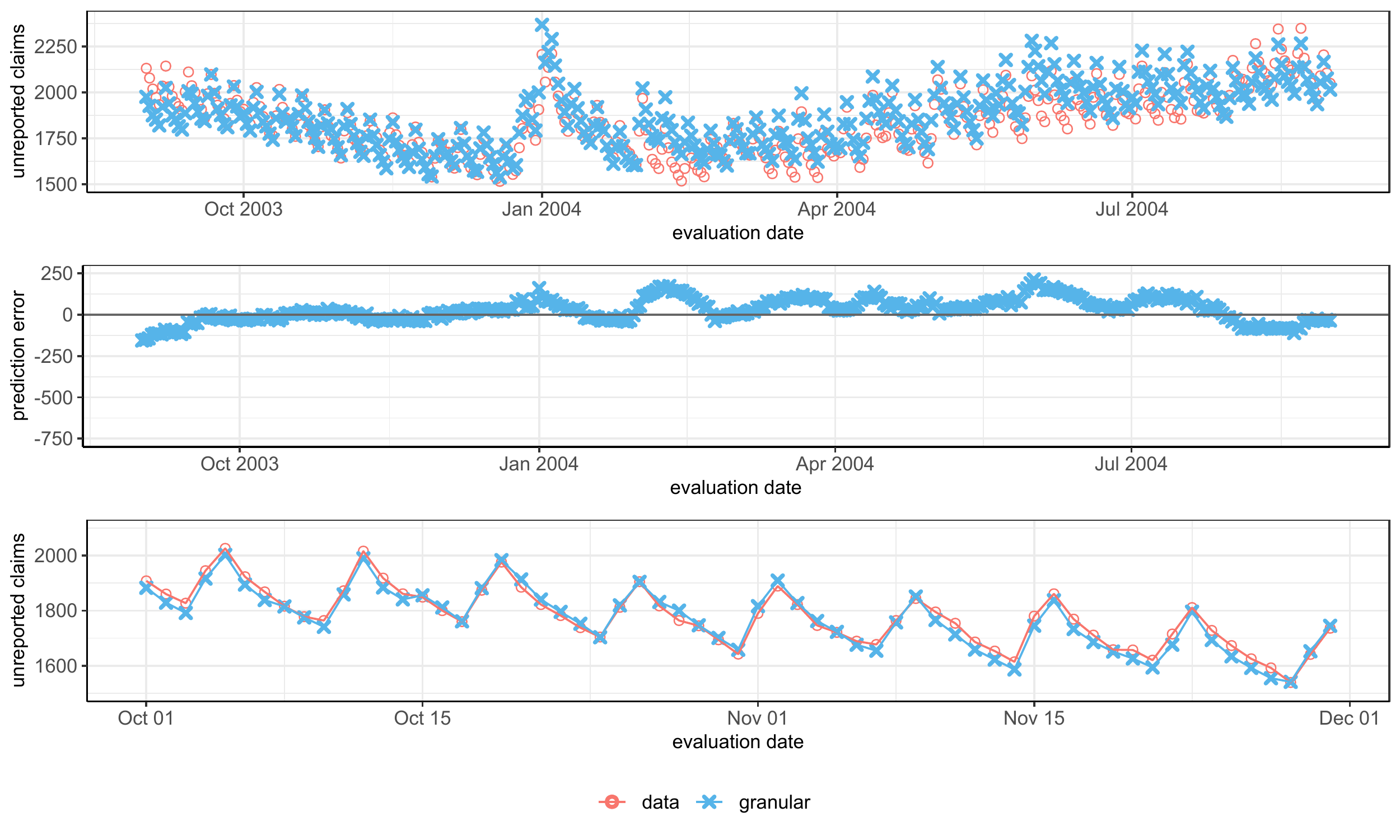}
\caption{Out-of-time prediction of the total IBNR count by the granular reserving method for each evaluation date between September 2003 and August, 2004. These estimates are compared with the observed values using data until August, 2009. The middle panel shows the difference between the predicted and actual IBNR count. The bottom panel zooms in on the estimates in October and November, 2003.}
\label{figure:totalIBNREst}
\end{figure}

\paragraph{Evolution of the number of hidden events} The primary focus of our granular model is estimating the total IBNR count. The top panel of Figure~\ref{figure:totalIBNREst} plots the predicted number of unreported claims on each evaluation date between September, 2003 and August, 2004. Each point estimate is an out-of-time IBNR estimate obtained from the granular model calibrated on the historical data available five days after the corresponding evaluation date. We compare these estimates with the actual number of IBNR claims computed from the data until August 31, 2009. Our model recognizes the trend in IBNR counts with more unreported claims during the summer compared to the winter months. The model also correctly predicts an increase in IBNR claims at the start of the year (here: January 1, 2004) as a result of the holidays in this period. The middle panel of Figure~\ref{figure:totalIBNREst} shows the prediction error, i.e.~the difference between the predicted number of IBNR claims and the actual count. The prediction error for the granular model is centred around zero and there are no large outliers. The bottom panel of Figure~\ref{figure:totalIBNREst} zooms in on the estimates for dates in October and November, 2003. This figure shows that the day of the week parameters allow the model to accurately capture the weekday pattern in IBNR counts.

\begin{figure}
\centering
\captionsetup[subfigure]{labelformat=empty}
\includegraphics[width = \textwidth]{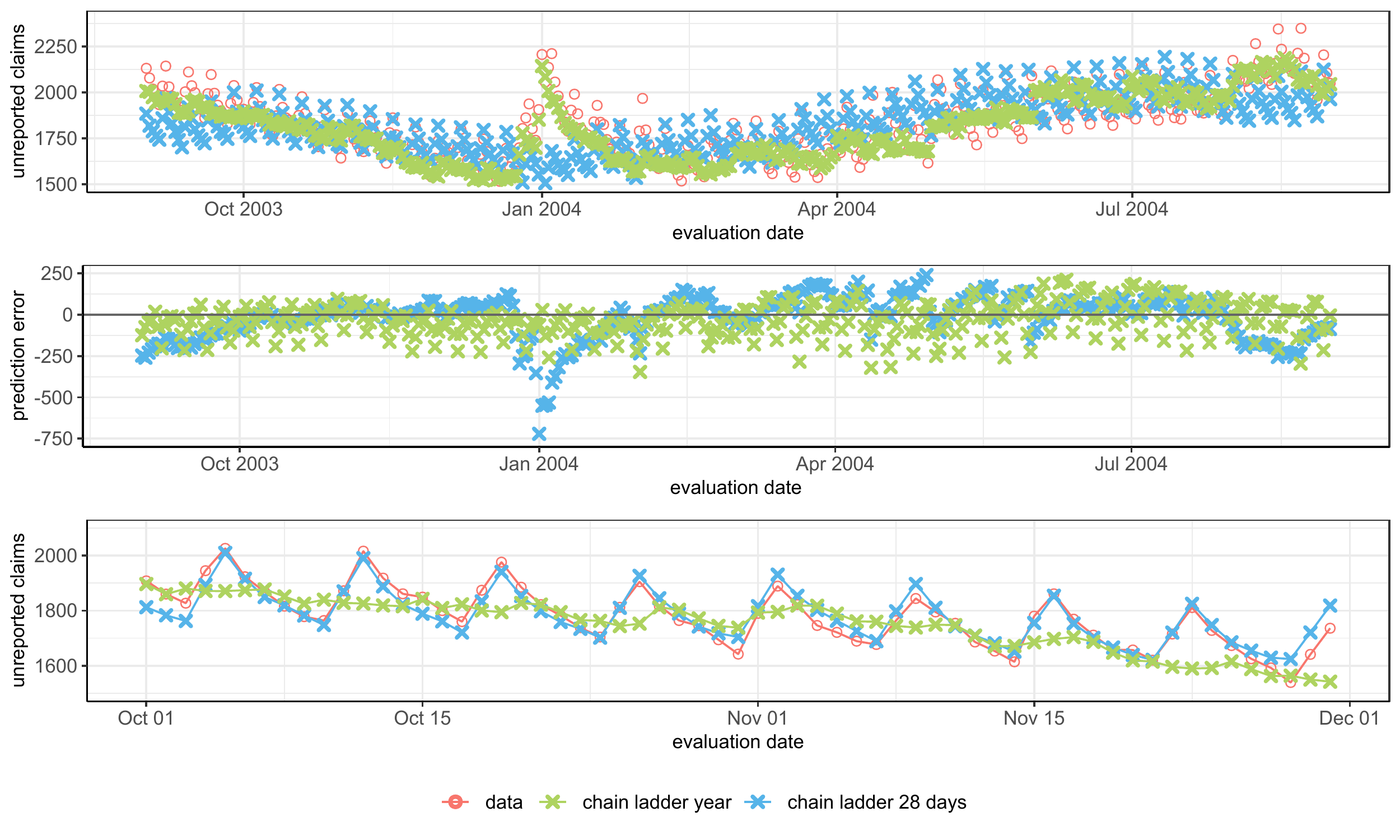}
\caption{Out-of-time prediction of the total IBNR count by the yearly and 28 day chain ladder methods for each evaluation date between September 2003 and August, 2004. These estimates are compared with the observed values using data until August, 2009. The middle panel shows the difference between the predicted and actual IBNR count. The bottom panel zooms in on the estimates in October and November, 2003.}
\label{figure:totalIBNREstCL}
\end{figure}

\paragraph{Benchmark with a model for aggregate data} We benchmark our granular approach to Mack's chain ladder method \cite{Mack1993} on aggregated data, which is the industry standard in claims reserving. This method discretizes time and aggregates the observed events into a two dimensional table based on the occurrence period and the discretized reporting delay. A Poisson generalized linear model (GLM) then models the effect of the occurrence and reporting period on these aggregated records. We investigate two aggregation levels, namely aggregating based on a yearly as well as a 28 day grid. We refer to \cite{Huang2015} for a more detailed discussion on reserving with granular data versus data aggregated in two dimensional tables. Figure~\ref{figure:totalIBNREstCL} shows the estimated IBNR counts under both chain ladder implementations evaluated on each date between September, 2003 and August, 2004. Both versions of the chain ladder detect the seasonal pattern in unreported claim counts, which is related to seasonality in the occurrence process. The end of the year holidays and corresponding increase in IBNR counts is a yearly seasonal effect in the reporting process. The chain ladder assumptions allow for seasonal effects when the period of seasonality coincides with the discretized time periods. For this reason, the yearly chain ladder method correctly predicts an increase in IBNR counts around the end of the year, whereas the 28 day chain ladder method severely underestimates IBNR counts for these dates. The bottom panel of Figure~\ref{figure:totalIBNREstCL} zooms in on the period October to November 2003. The 28 day chain ladder method retrieves the day of the week effect, since the length of every bin is a multiple of 7 and therefore contains the same weekdays. The yearly chain ladder method has bins with either 365 or 366 days. Since both bin sizes are not divisible by 7, the yearly chain ladder method is unable to recognize the day of the week effect. This results in a systematic overestimation of IBNR counts on Fridays and an underestimation on Sunday. The middle panel of Figure~\ref{figure:totalIBNREstCL} shows the difference between the predicted and actual IBNR count. The inability of the 28 day chain ladder to capture the holiday effect results in large underestimations around this time of the year. The yearly chain ladder overall performs better, but the prediction error is sensitive to the day of the week on which the reserve is calculated. Capturing the holiday and the day of the week effect simultaneously requires a model specified at the daily level. The chain ladder method assumes independence between the reporting delay distribution and the occurrence period of the claim. Since Figure~\ref{figure:CalendarEffect} and \ref{figure:delayDistributionThreeWeeks} indicate that this assumption is not valid at the daily level, a daily chain ladder would not perform well. Our granular method explains both phenomena together by abandoning this independence assumption.

\subsection{Scenario testing} \label{section:simulation}

\subsubsection{Investigated scenarios} \label{section:scenarios}
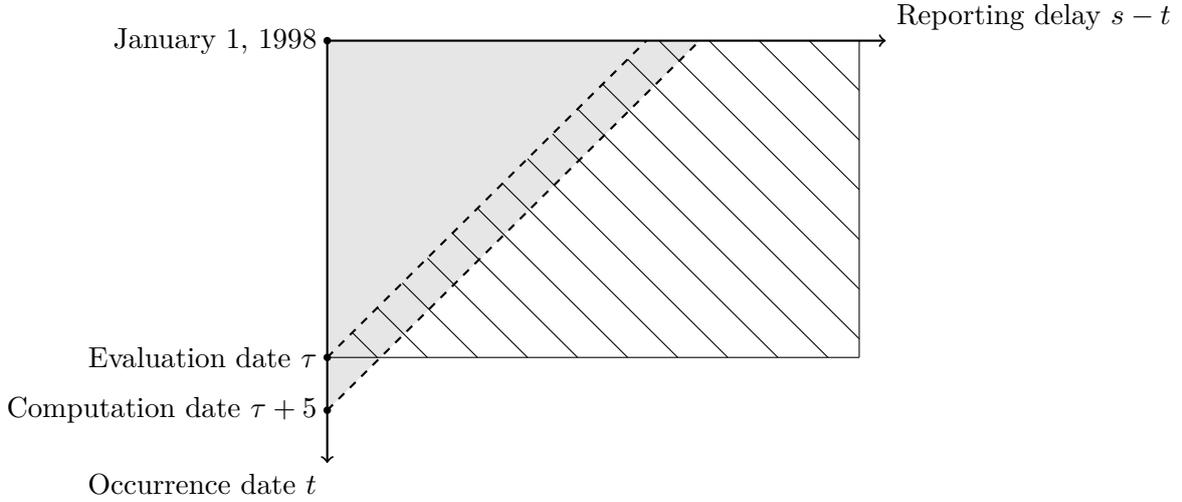
\begin{figure}[ht!]
\begin{tikzpicture}[scale = 0.7]
\tikzset{hstyle/.style={solid,thin}}


\fill [gray!20](5, 9) -- (5, 2) -- (12, 9) -- cycle;

\def\n{17}
\foreach \i in {1,...,\n}
{
	\pgfmathsetmacro{\xa}{5 + \i * 8 / \n};
	\pgfmathsetmacro{\xb}{5 + \i * 16 / \n};
	\pgfmathsetmacro{\y}{\i * 8 / \n + 3};
	
	\pgfmathsetmacro{\da}{min(9 - \y, 0)};
	\pgfmathsetmacro{\db}{min(15 - \xb, 0)};

    \draw (\xa - \da, \y + \da) -- (\xb + \db, \y - \xb + \xa - \db);
);
}

\draw (15, 9) -- (15, 3);
\draw (5, 3) -- (15, 3);

\draw[->, thick] (5,9) -- (15.5, 9) node[right, anchor = south west]{Reporting delay $s-t$};
\draw[->, thick] (5,9) -- (5, 1) node[above, anchor = north east]{Occurrence date $t$};

\draw [dashed, thick](5, 3) -- (11, 9);
\draw [dashed, thick](5, 2) -- (12, 9);

\node[fill,circle, inner sep=1pt] at (5, 9) {};
\node at (5, 9) [left] {January 1, 1998};

\node[fill,circle, inner sep=1pt] at (5, 3) {};
\node at (5, 3) [left] {Evaluation date $\tau$};

\node[fill,circle, inner sep=1pt] at (5, 2) {};
\node at (5, 2) [left] {Computation date $\tau + 5$};

\end{tikzpicture}
\caption{Structure of a simulated data set. We simulate accidents that occur between the first of January, 1998 and the computation date, together with their associated reporting delay. The gray area shows the data that is used to fit the model and to predict the hatched area, which consists of the number of unreported claims at the evaluation date $\tau$. We obtain perfect predictions for the intersection of the gray area and the hatched area, since in this region the reported counts are observed.}
\label{figure:simulationOverview}
\end{figure}

\begin{landscape}
\begin{figure}
\centering
\includegraphics[width=0.95\linewidth]{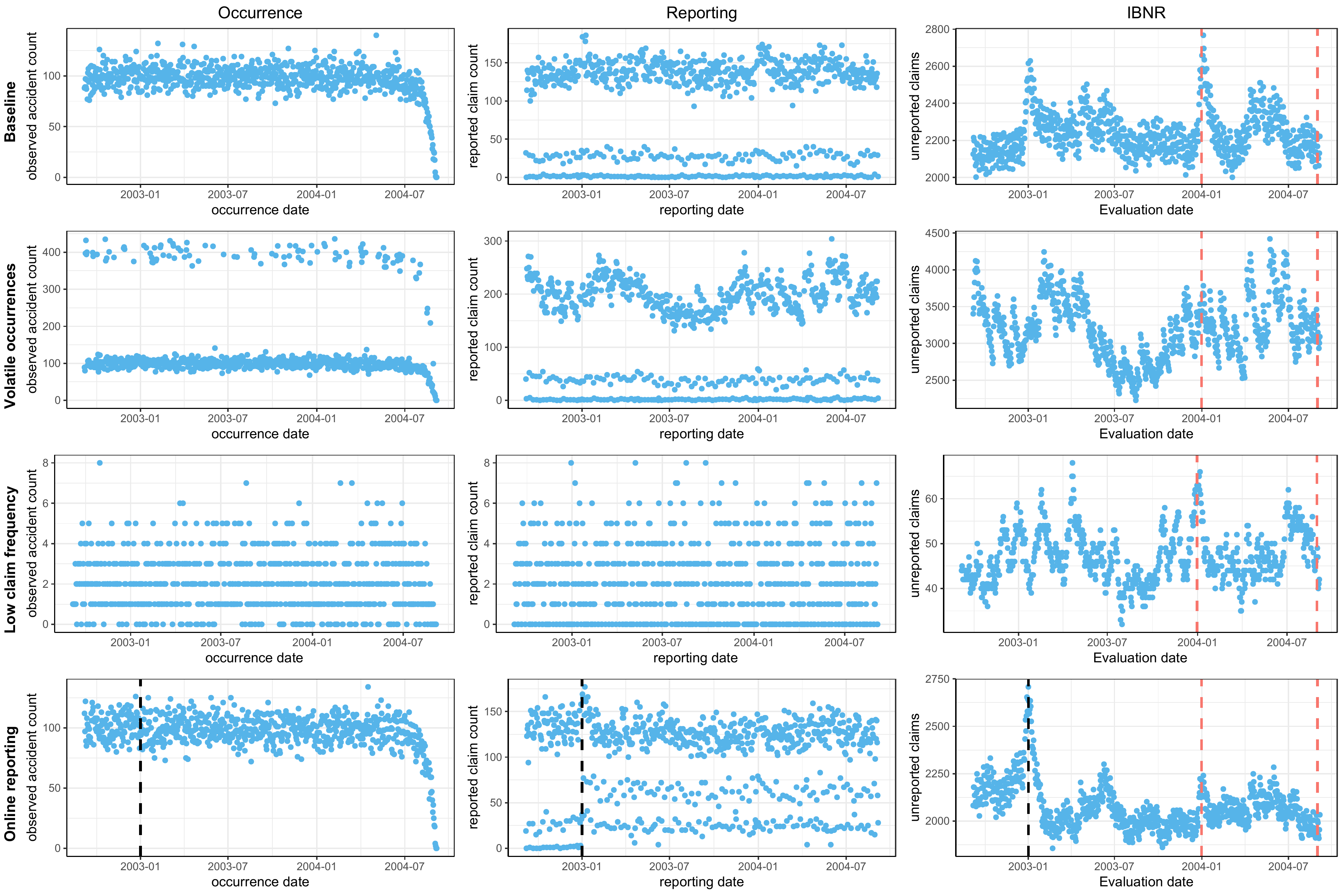}
\captionsetup{width=0.95\linewidth}
\caption{ Each row visualizes a simulated data set from one of the four scenarios. The left column shows the daily number of accidents that were reported by August 31, 2004 (cf.~Figure~\ref{figure:ClaimCount}). The middle column shows the daily number of reported claims (cf.~Figure~\ref{figure:ReportCount}). The right column visualizes the number of unreported accidents using a rolling evaluation date (cf.~Figure~\ref{figure:totalIBNR}). The red dashed lines in the IBNR plots indicate the evaluation dates of December 31, 2003 and August 31, 2004.}
\label{figure:scenarioGrid}
\end{figure}
\end{landscape}

We further evaluate our approach with portfolios simulated along four different scenarios. Each scenario generates data from an insurance portfolio from January 1, 1998 onwards. Figure~\ref{figure:simulationOverview} outlines the structure of these data sets. The insurer observes the claims that are reported before the computation date (the gray area in Figure~\ref{figure:simulationOverview}) and predicts the number of claims that were not yet reported on the evaluation date (the hatched area in Figure~\ref{figure:simulationOverview}). We consider two evaluation dates (December 31, 2003 and August 31, 2004) to visualize the impact of holidays near the end of the year on the accuracy of IBNR claim count predictions. The four scenarios focus on characteristics of the portfolio or the claim handling process that have an impact on the total IBNR count. Figure~\ref{figure:scenarioGrid} visualizes the occurrence, reporting and IBNR processes for a single simulated data set from each of the four scenarios.

\paragraph{Scenario 1: Baseline scenario}
This is the basic scenario from which the other three scenarios will slightly deviate. The occurrence of insured events follows a Poisson distribution with an average of 100 claims on each occurrence date. For these occurrences the reporting delay is simulated along the model specification outlined in Section~\ref{section:model}, i.e.~the distribution of the time-changed reporting delay $\tilde{U}$ follows a lognormal distribution with density
\begin{equation*}
	f_{\tilde{U}}(u) = \frac{1}{u \sigma \sqrt{2\pi}} e^{- \frac{1}{2} \cdot \left(\frac{\ln(u) - \mu}{\sigma} \right)^2},
\end{equation*}
where $\mu = 0$ and $\sigma = 1$. The daily reporting \weight{} depends only on the reporting date and is given by
\begin{equation*}
	\alpha_{t, s} = 0.10 \cdot (0.20)^{\characteristic{s \in \texttt{Sat}} + \characteristic{s \in \texttt{unofficial-holiday}}} \cdot (0.01)^{\characteristic{s \in \texttt{Sun}} + \characteristic{s \in \texttt{national-holiday}}},
\end{equation*}
where \texttt{Sat}, \texttt{Sun}, \texttt{national-holiday} and \texttt{unofficial-holiday} are the sets of all Saturdays, Sundays, national holidays and unofficial holidays respectively. As such, the reporting probability is reduced by $80\%$ on Saturdays and unofficial holidays and by $99\%$ on Sundays and national holidays. These effects are of the same order as those found in the exploratory data analysis, see e.g.~Figure~\ref{figure:CalendarEffect} in Section~\ref{section:dataDescription} and result in an average reporting delay of slightly more than three weeks. The top row of Figure~\ref{figure:scenarioGrid} visualizes a simulation from this baseline scenario. The middle panel shows two regimes of reporting, where the days with few reported claims correspond to the weekend and holidays.

\paragraph{Scenario 2: Volatile occurrences}
In this scenario external causes, such as the weather, have a large effect on the number of accidents that occur on a given date. The environment can be in two states, a good state with an average of 100 accidents per day and a bad state in which there are on average 400 accidents.  The transitions between these states follow a Markov process with transition matrix
$$
\bordermatrix{\text{from/to} & \text{good} & \text{bad} \cr
                  \text{good} & 0.9 & 0.1 \cr
                  \text{bad} & 0.6 & 0.4 \cr}.
$$
The model starts in the good state and then occasionally moves to the bad state. From this bad state there is a large probability of returning to the good state with less occurrences on average. The second row of Figure~\ref{figure:scenarioGrid} (lhs) visualizes the impact of this bad state on the occurrence process.  The reporting delay distribution is the one described in the baseline scenario.

\paragraph{Scenario 3: Low claim frequency}
This scenario illustrates the effect of a strong reduction in the number of occurred accidents. The occurrence process is modeled by a Poisson distribution with a daily average of two claims. The reporting model from the baseline scenario is used. This scenario is visualized in the bottom row of Figure~\ref{figure:scenarioGrid}. We observe that a low number of accidents leads to more volatility in the IBNR process.

\paragraph{Scenario 4: Online reporting}
In this scenario the insurer introduces an online tool for claim reporting. This online tool is launched at January 1, 2003 and increases the number of reports in the weekend and on holidays. The new reporting \weights{} become
\begin{equation*}
	\alpha_{t, s} =
	\begin{cases}
		0.10 \cdot (0.20)^{\characteristic{s \in \texttt{Sat}} + \characteristic{s \in \texttt{Unofficial-holiday}}} \cdot (0.01)^{\characteristic{s \in \texttt{Sun}} + \characteristic{s \in \texttt{Holiday}}} & s < 01/01/2003 \\
		0.10 \cdot (0.50)^{\characteristic{s \in \texttt{Sat}} + \characteristic{s \in \texttt{Unofficial-holiday}}} \cdot (0.20)^{\characteristic{s \in \texttt{Sun}} + \characteristic{s \in \texttt{Holiday}}} & s \geq 01/01/2003
	\end{cases}.
\end{equation*}
This reporting model is combined with the same occurrence process as in the baseline model, that is a Poisson process with a constant intensity of 100 claims each day. The bottom row of Figure~\ref{figure:scenarioGrid} visualizes a simulation from this scenario. A vertical black line indicates the breakpoint on January 1, 2003. After the introduction of online reporting we no longer observe dates with zero reports. 

\subsubsection{Calibrated models: granular versus aggregate} \label{section:modeldesign}

We compare the accuracy of the predictions of the hidden event counts using three models, namely the exact granular model from which we simulated the data, an approximate granular model and a model for yearly aggregated data. The historical information (gray area in Figure~\ref{figure:simulationOverview}) is used to predict the number of IBNR claims (hatched area in Figure~\ref{figure:simulationOverview}). Under the granular approach these predictions naturally extend to delays beyond those yet observed, whereas in the aggregate approach we limit the prediction window to the longest observed delay. We consider a gap of five days between the computation and the valuation date. The observations from these five days improve the prediction of the occurrence intensities $\lambda_t$ and the reporting probabilities $p_{t, s}$, whereas there is no straightforward way to incorporate this data in the method for yearly, aggregated data. The ability to use this additional data is one of the advantages of the granular approach.

\paragraph{Exact granular model} We use our knowledge of the shape of the distribution and reporting exposure structure behind the various scenarios and calibrate the exact same model for reporting delay on the historical data. Hence we estimate the variance parameter in the lognormal distribution for the smoothed reporting delay $\tilde{U}$ and the parameters $\vectorF{\gamma}$ for the covariate effects in the reporting exposures $\alpha_{t, s}$. The reporting exposure $\alpha_{t, s}$ changes the scale of the time axis which is similar to the effect of the scale parameter $\exp(\mu)$ of the lognormal distribution. We avoid identifiability issues by setting $\mu$ equal to zero. The occurrence process is modeled non-parametrically as described in Section~\ref{section:model}.

\paragraph{Approximate granular model} This model considers the more realistic situation where the insurer wants to fit the model of Section~\ref{section:model}, but is unaware of the exact underlying distribution. Motivated by computational benefits the insurer chooses an exponential distribution for the smoothed reporting delay $\tilde{U}$, and structures the reporting exposures as
\begin{align} \label{eq:reportingExposureApproximate}
	\alpha_{t, s} &= \alpha_{s}^{\texttt{dow}} \cdot \alpha_{s}^{\texttt{holiday}} \cdot \alpha^{\texttt{delay}}_{s-t} \\
					&= \exp(
\covariateDisplay{(\vectorF{x}_{s}^{\texttt{dow}})^{'} \cdot \gamma^{\texttt{dow}}}{\vectorF{\gamma}^{\texttt{dow}} \cdot \texttt{dow(s)}}  +
\covariateDisplay{(\vectorF{x}_{s}^{\texttt{holiday}})^{'} \cdot \gamma^{\texttt{holiday}}}{\vectorF{\gamma}^{\texttt{holiday}} \cdot \texttt{holiday(s)}} +
\covariateDisplay{(\vectorF{x}_{s-t}^{\texttt{delay}})^{'} \cdot \gamma^{\texttt{delay}}}{\vectorF{\gamma}^{\texttt{delay}} \cdot \texttt{delay(s-t)}} ). \nonumber
\end{align}
In this specification $\alpha_{s}^{\texttt{dow}}$ captures the day of the week effect, $\alpha_{s}^{\texttt{holiday}}$ identifies national and unofficial holidays and $\alpha_{s-t}^{\texttt{delay}}$ adapts reporting exposure based on the time elapsed since the claim occurred. For a single simulated data set we bin reporting delay in 13 bins according to the strategy outlined in online appendix~\ref{section:AdaptingExponential}. These same bins are then reused to construct the delay covariate for all other simulations. In the fourth scenario (online reporting), we estimate different parameter values for the parameters $\gamma^{\texttt{dow}}$ and $\gamma^{\texttt{holiday}}$ for reporting dates before and after January 1, 2003.

\paragraph{A model for aggregated data: the chain ladder} The chain ladder method described in Section~\ref{section:modelPredictions} is the industry standard for predicting the number of unreported claims. We aggregate the simulated data by calendar year and benchmark our granular approach to the chain ladder method on this aggregated data.

\subsubsection{Results and discussion}
\begin{figure}
\hfill
\includegraphics[width=0.95\textwidth]{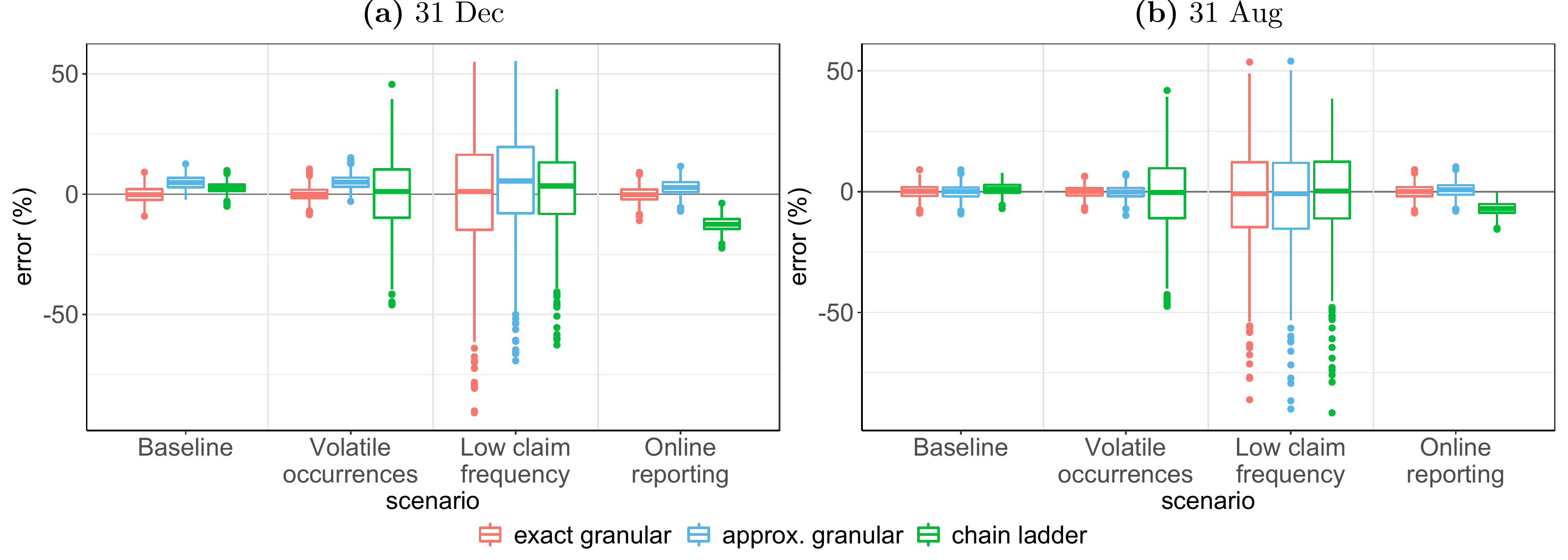}
\hfill
\caption{Boxplots of the Percentage Error (PE) of the IBNR estimate across the four scenarios and on both evaluation dates.}
\label{figure:boxplotSimulation}
\end{figure}

We evaluate the performance of the reserving models by predicting the total number of IBNR claims at the evaluation date, which corresponds to the hatched area in Figure~\ref{figure:simulationOverview}. This prediction is compared with the actual number of unreported claims as observed in the simulated data set. We simulate \num{1000}  data sets and calibrate the three models outlined in Section~\ref{section:modeldesign} on each of these. The prediction accuracy is measured by the percentage error (PE), i.e.
\begin{align*}
	\text{PE} = 100 \cdot \frac{N^{\IBNR}(\tau) - \widehat{N^{\IBNR}(\tau)}}{N^{\IBNR}(\tau)}.
\end{align*}
Positive percentage errors reflect underestimation, whereas negative values indicate an overestimation of IBNR counts. Table~\ref{table:simulationOverview} shows the mean and standard deviation of the percentage error for the two granular models and the chain ladder method. In Figure~\ref{figure:boxplotSimulation} boxplots of the percentage error visualize the model performance across the four scenarios.

\begin{table}
\centering
\begin{tabular}{ l  l  l l  l l l l} \toprule
\multirow{2}{*}{Scenario} & \multirow{2}{*}{Eval. date}  &  \multicolumn{2}{c}{exact granular} & \multicolumn{2}{c}{approx. granular} & \multicolumn{2}{c}{chain ladder} \\
& & $\mu(PE)$ & $\sigma(PE)$ & $\mu(PE)$ & $\sigma(PE)$ & $\mu(PE)$ & $\sigma(PE)$  \\ \midrule
\multirow{2}{*}{Baseline} &  31 Dec 2003  & -0.09 & 3.17 & 4.85 & 2.75 & 2.70 & 2.17\\
&  31 Aug 2004  & -0.01 & 2.75 & -0.18 & 2.82 & 1.20 & 2.36  \\ \midrule
\multirow{2}{*}{Volatile occurrences} &  31 Dec 2003  & 0.11 & 2.64 & 5.01 & 2.93 & 0.16 & 15.52 \\
&  31 Aug 2004 & -0.04 & 2.27 & -0.20 & 2.51 & -0.82 & 14.90 \\ \midrule
\multirow{2}{*}{Low claim frequency} &  31 Dec 2003  & -0.69 & 23.89 & 4.42 & 20.85 & 1.65 & 16.25\\
&  31 Aug 2004  &  -2.30 & 20.19 & -2.52 & 20.72 & -1.33 & 17.96 \\ \midrule
\multirow{2}{*}{Online reporting} & 31 Dec 2003  & -0.13 & 3.12 & 2.93 & 3.07 & -12.46 & 2.91 \\
&  31 Aug 2004  & 0.02 & 2.80 & 0.73 & 2.89 & -7.00 & 2.68   \\ \bottomrule
\end{tabular}
\caption{Evaluation of the performance of the exact granular model, the approximate granular model and the chain ladder method across four different scenarios and two evaluation dates.}
\label{table:simulationOverview}
\end{table}

\paragraph{Impact of evaluation date} We observe in all four scenarios an increase in unreported claims on New Year's Eve (see the last column in Figure~\ref{figure:scenarioGrid}). This is the result of multiple holidays at the end of the year, which prevents clients from reporting their claim. We compare the average percentage error in Table~\ref{table:simulationOverview} on December 31, 2003 and August 31, 2004 to quantify the impact of these holidays on prediction accuracy. The exact granular model fits the distributional specification that was used in the simulation. Therefore this model can perfectly capture the effect of holidays and has an average error close to zero on both dates. Seasonal effects do not violate the chain ladder assumptions when their seasonal cycle coincides with the chain ladder period. Since the end of the year holidays can be seen as a yearly seasonal event they do not affect the prediction accuracy in the yearly chain ladder method. This explains the fairly similar errors on both evaluation dates for the chain ladder method. Table~\ref{table:simulationOverview} reveals an underestimation of IBNR counts for the approximate granular model on December 31 across all four scenarios. The data is simulated with a lognormal distribution for the smoothed reporting delay, whereas in the approximate granular model we fit an exponential distribution. Since these distributions are quite different, we include a delay effect $\alpha_{s-t}^{\texttt{delay}}$ in $\eqref{eq:reportingExposureApproximate}$. This effect can increase the reporting probability at specific delays, hereby moving the time-changed data closer to an exponential distribution. However, the delay covariate can not remove all differences between these distributions and this leads to a small underestimation on December 31, 2004 in all scenarios. For all three models the choice of evaluation date does not influence the standard deviation of the percentage error.

\paragraph{Baseline} The top row of Figure~\ref{figure:scenarioGrid} visualizes a single data set from the baseline scenario. Both the occurrence and reporting process are stable. This leads to a yearly periodical pattern in IBNR counts, which is easy to predict. Since all three models perform well (see Figure~\ref{figure:boxplotSimulation}), there is no reason to replace the chain ladder method by a granular model in this scenario.

\paragraph{Volatile occurrences} The range of IBNR values encountered throughout a year is much wider in this scenario compared to the other three scenarios. Table~\ref{table:simulationOverview} and Figure~\ref{figure:boxplotSimulation} show that the performance of the granular models is in line with their performance in the baseline scenario. The occurrence process has little effect on the prediction accuracy, since we model the occurrence process non-parametrically. The chain ladder method performs well on average, but the standard deviation has risen compared to the baseline scenario. In over half of the cases the chain ladder produces an error of more than $10\%$ when predicting the number of unreported claims. The chain ladder method aggregates claims by occurrence year, hereby losing the exact occurrence information. When the model was in the bad state on the evaluation date, this leads to large underestimations of total IBNR counts. This scenario identifies an unstable accident occurrence process as a reason for considering a granular model.

\paragraph{Low claim frequency} The occurrence frequency is reduced from an average of hundred daily claims to only two claims. The third row of Figure~\ref{figure:scenarioGrid} visualizes a data set from this scenario. Since on average only two accidents occur per day, our predictions for the intensities $\lambda_t$ in the occurrence process are less reliable. As seen in Figure~\ref{figure:boxplotSimulation} this leads to large prediction errors for all models. This uncertainty follows mostly from the Poisson assumption (A1) in the data generation process. The coefficient of variation $\frac{\sigma}{\mu}$ for a Poisson distribution with intensity $\lambda$ is given by $\frac{1}{\sqrt{\lambda}}$. A lower intensity in the Poisson proccess corresponds with a larger coefficient of variation and thus more uncertainty in the data. We conclude that accurate estimation of the number of hidden events is only possible when the expected number of events is sufficiently large.

\paragraph{Online reporting} On January 1, 2003 the insurer introduces an online tool to report claims, which creates a breakpoint in the reporting process. 
The granular model performs well on both evaluation dates, since we estimate different exposure parameters after the breakpoint. Both evaluation dates correspond with around one year of post breakpoint data, which is insufficient for applying the chain ladder method. Therefore, we calibrate the chain ladder method on all the available data, which leads to an overestimation of the IBNR counts. This scenario illustrates the benefits of a granular reserving model, when breakpoints can be identified in the data.

\section{Conclusion} \label{section:conclusion}
We propose a new method to model the number of events that occurred in the past, but which are not yet registered due to an observation delay. Our approach provides an elegant and flexible framework for modeling the observation delay subject to calendar day covariates by introducing the concept of observation exposure. This framework can be applied for predicting the future cost of warranties, pricing maintenance contracts and many other applications in operational research where events are observed with a delay.  We illustrate our method in an extensive insurance case-study. Compared to methods designed for aggregated data our granular approach has three advantages. First of all, introducing covariates gives insight into the observation process. Second, our granular model can predict the expected number of observations for each future date. This enables the detection of changes in the reporting process in a fast way. Third, by introducing covariates the predictive performance is less sensitive to the chosen evaluation date. The simulation study further identifies a volatile occurrence process and breakpoints in the event observation process as important arguments for choosing a data driven, granular model as developed in this paper.

\section{Acknowledgments}
The authors thank the anonymous referees and the associate editor for useful comments which led to significant improvements of the paper. This work was supported by the Argenta Research chair at KU Leuven; KU Leuven's research council [project COMPACT C24/15/001]; the agency for Innovation by Science and Technology (IWT) [grant number 131173]; and Research Foundation Flanders (FWO) [grant number 11G4619N]. 

\bibliography{references} 

\appendix

\title{Supplementary material for \\ "Modeling the number of hidden events subject to observation delay"}
\date{\today}
\maketitle

\appendix 

\section{Maximum likelihood estimation of observation exposure parameters} \label{appendix:fittingProcedure}
We model a parameter vector $\vectorF{\gamma}$ which structures the observation exposures. 
\begin{align}
	\ell(\vectorF{\gamma} ; \vectorF{\chi} ) &= \sum_{t=1}^\tau \sum_{s=t}^\tau N_{t, s} \cdot \log (p_{t, s} ) - \sum_{t=1}^\tau N_t^{\Rep}(\tau) \cdot \log(p_t^{\Rep}(\tau) )\label{eq:loglikelihoodTime}\\
& = \sum_{t = 1}^\tau \sum_{s = t}^\tau N_{t, s} \cdot \log\left(F_{\tilde{U}}\left(\varphi_{t}(s-t+1) \right) - F_{\tilde{U}} \left(\varphi_{t}(s-t) \right) \right) \nonumber \\
 &\phantom{ = }-\sum_{t=1}^\tau N_t^{\Rep}(\tau) \cdot \log\left(F_{\tilde{U}}\left(\varphi_{t}(\tau-t+1) \right)\right), \nonumber
\end{align}
where
\begin{align*}
	\varphi_{t}(d) = \sum_{v=t}^{t+d-1} \exp(\vectorF{x}_{t, v}^{'} \vectorF{\gamma}).
\end{align*}
No analytical solution exists for the optimal parameters $\vectorF{\gamma}$ and numerical optimization is required. We use the Newton-Raphson algorithm to maximize the likelihood \eqref{eq:loglikelihoodTime}. The Newton-Raphson algorithm updates the parameter estimates iteratively as follows
\begin{equation}
	\hat{\vectorF{\gamma}}^{(k+1)} = \hat{\vectorF{\gamma}}^{(k)} - \vectorF{H}^{-1}(\hat{\vectorF{\gamma}}^{(k)}) \cdot \vectorF{S}(\hat{\vectorF{\gamma}}^{(k)}). \label{eq:iterativeNR}
\end{equation}
In this formula $\vectorF{S}$ denotes the score vector and $\vectorF{H}$ is the Hessian of the loglikelihood in $\eqref{eq:loglikelihoodTime}$, i.e.~the vector of first order and the matrix of second order partial derivatives respectively. Below we derive the expression for the first and second order derivatives of the loglikelihood when $F_{\tilde{U}}$ is a known twice continuously differentiable distribution function. The components of the score vector $\vectorF{S}$ are
\begin{small}
\begin{align}
	\frac{\partial \ell(\vectorF{\gamma}, \vectorF{\xi} ; \vectorF{\chi})}{\partial \gamma_i} = &
		\sum_{t=1}^\tau \sum_{s=t}^\tau  \frac{N_{t, s}}{p_{t,s}} \cdot \left[ 
			f_{\tilde{U}}\left( \varphi_{t}(s-t+1) \right) \cdot \partialD{\varphi_t}{\gamma_i}(s-t+1)- 
			f_{\tilde{U}}\left( \varphi_{t}(s-t) \right) \cdot \partialD{\varphi_t}{\gamma_i}(s-t) \right]
  \nonumber\\
	& - \sum_{t=1}^\tau \frac{N_t^{\Rep}(\tau)}{p_t^{\Rep}(\tau)} \cdot 
		f_{\tilde{U}}\left( \varphi_{t}(\tau-t+1) \right) \cdot 
		\partialD{\varphi_t}{\gamma_i}(\tau - t+1), \nonumber
\end{align}
\end{small}
where $f_{\tilde{U}}(\, \cdot \,)$ denotes the density function of $F_{\tilde{U}}(\, \cdot \,)$ and
\begin{align*}
	p_{t, s} &= F_{\tilde{U}}\left(\varphi_{t}(s-t+1) \right) - F_{\tilde{U}} \left(\varphi_{t}(s-t) \right) \\
	p_{t, s}^{\Rep}(\tau) &= F_{\tilde{U}}\left(\varphi_{t}(\tau-t+1) \right).
\end{align*}
The derivatives of the time change operator $\varphi_t$ with respect to $\vectorF{\gamma}$ are
\begin{align*}
	\frac{\partial}{\partial \gamma_i} \varphi_{t}(s-t+1) = \sum_{v=t}^{s} x_{t, v, i} \cdot \alpha_{t, v}
\end{align*}
where $x_{t, s, i}$ is the covariate value of the $i$-th parameter for reporting on date $s$ for a claim that occurred on date $t$. The Hessian $\vectorF{H}$ is given by
\begin{small}
\begin{align*}
	& \partialDD{\ell(\vectorF{\gamma} ; \vectorF{\chi})}{\gamma_i}{\gamma_j} = \nonumber \\
	& \qquad \sum_{t=1}^{\tau} \sum_{s=t}^\tau  \frac{N_{t, s}}{p_{t, s}} \cdot \Bigg[ 
		f_{\tilde{U}}^{'}\left(\varphi_{t}(s-t+1) \right) \cdot 
		\partialD{\varphi_t}{\gamma_i} (s-t+1)\cdot 
		\partialD{\varphi_t}{\gamma_j} (s-t+1) \\
	& \qquad \qquad - 
		f_{\tilde{U}}^{'}\left(\varphi_{t}(s-t) \right) \cdot 
		\partialD{\varphi_t}{\gamma_i} (s-t)\cdot 
		\partialD{\varphi_t}{\gamma_j} (s-t) \nonumber \\
    & \qquad \qquad +
		f_{\tilde{U}}\left( \varphi_{t}(s-t+1) \right) \cdot 
		\partialDD{\varphi_t}{\gamma_i}{\gamma_j}(s-t+1) - 
		f_{\tilde{U}}\left( \varphi_{t}(s-t) \right) \cdot 
		\partialDD{\varphi_t}{\gamma_i}{\gamma_j}(s-t) \Bigg] \nonumber \\  	
	& \qquad - \sum_{t=1}^{\tau} \sum_{s=t}^\tau  \frac{N_{t, s}}{p_{t, s}^2} \cdot \Bigg[ 
		f_{\tilde{U}}\left( \varphi_{t}(s-t+1) \right)^2 \cdot 
		\partialD{\varphi_t}{\gamma_i} (s-t+1)\cdot 
		\partialD{\varphi_t}{\gamma_j} (s-t+1)  \\
	& \qquad \qquad 
		+ f_{\tilde{U}}\left( \varphi_{t}(s-t) \right)^2 \cdot 
		\partialD{\varphi_t}{\gamma_i} (s-t) \cdot 
		\partialD{\varphi_t}{\gamma_j} (s-t) \\
	& \qquad \qquad - 
		f_{\tilde{U}}\left( \varphi_{t}(s-t+1) \right) \cdot 
		f_{\tilde{U}}\left( \varphi_{t}(s-t) \right) \cdot 
		\partialD{\varphi_t}{\gamma_i}(s-t+1) \cdot 
		\partialD{\varphi_t}{\gamma_j}(s-t) \\
	& \qquad \qquad - 
		f_{\tilde{U}}\left( \varphi_{t}(s-t+1) \right) \cdot 
		f_{\tilde{U}}\left( \varphi_{t}(s-t) \right) \cdot 
		\partialD{\varphi_t}{\gamma_i}(s-t) \cdot 
		\partialD{\varphi_t}{\gamma_j}(s-t+1) 
		\Bigg] \\
	& \qquad - \sum_{t=1}^{\tau} \frac{N_t^{\Rep}(\tau)}{p_t^{R}(\tau)} \cdot \Bigg[
		f_{\tilde{U}}^{'}\left(\varphi_{t}(\tau - t+1) \right) \cdot 
		\partialD{\varphi_t}{\gamma_i} (\tau-t+1)  \cdot 
		\partialD{\varphi_t}{\gamma_j} (\tau-t+1) \\
	& \qquad \qquad + 
			f_{\tilde{U}}(\varphi_t(\tau-t+1)) \cdot 
			\partialDD{\varphi_t}{\gamma_i}{\gamma_j}(\tau - t+1) \Bigg] \\	
	& \qquad + \sum_{t=1}^{\tau} \frac{N_t^{\Rep}(\tau)}{p_t^{\Rep}(\tau)^2} \cdot  
			f_{\tilde{U}}\left(\varphi_{t}(\tau - t+1) \right)^2 \cdot 
			\partialD{\varphi_t}{\gamma_i}(\tau-t+1)  \cdot 
			\partialD{\varphi_t}{\gamma_j}(\tau-t+1), 
\end{align*}
\end{small}
where the second order derivatives of $\varphi_t$ with respect to $\vectorF{\gamma}$ are
\begin{align*}
	\frac{\partial}{\partial \gamma_i \partial \gamma_j} \varphi_{t}(s-t+1) = \sum_{v=t}^{s} x_{t, v, i} \cdot x_{t, v, j} \cdot \alpha_{t, v}
\end{align*}

The Newton-Raphson algorithm in \eqref{eq:iterativeNR} models the observation exposure parameters $\vectorF{\gamma}$. Together with the observation parameters, the simulation study of Section~\ref{section:simulation} estimates the variance parameter $\sigma$ in the lognormal time-changed distribution. The Newton-Raphson algorithm in $\eqref{eq:iterativeNR}$ can easily be extended to this case, where the distribution function of $F_{\tilde{U}}$ depends on parameters.

\section{Simulation procedure} \label{appendix:simulation}
We outline the algorithm that was used to generate data sets from the four scenarios specified in Section~\ref{section:scenarios}. This algorithm combines a model for the occurrence of events with a model for the observation delay as described in Section~\ref{section:model}. We divide the algorithm in three steps.
\paragraph{Step 1. Occurrence} We first generate the number of occurred events. The number of daily events follows a Poisson distribution
\begin{equation*}
	N_t \sim \text{Poisson}(\lambda_t),
\end{equation*}
where the intensity $\lambda_t$ is obtained from the occurrence process specification for the scenarios in Section~\ref{section:simulation}.

\paragraph{Step 2. Observation} We now simulate the observation date for each occurred event. Combining equation $\eqref{eq:transformation}$ and $\eqref{eq:formulaP}$, we can write the probability that an event from date $t$ is observed on date $s$ as
\begin{equation*}
	p_{t, s} = P\left(\tilde{U} \in \left[ \sum_{v=t}^{s-1} \alpha_{t, v}, \sum_{v=t}^{s} \alpha_{t, v} \right) \right). 
\end{equation*}
We define the observation date random variable
\begin{equation}
	S_t = \min_{s} \bigg\{s \in \mathbb{N} \,\Big| \, \sum_{v=t}^{s} \alpha_{t, v} > \tilde{U}\bigg\}. \label{eq:simulReportingDate}
\end{equation}
This expression transforms the time-changed observation delay random variable into the associated observation date. Consequently $S_t$ satisfies $P(S_t = s) = p_{t, s}$. For each event that occurred on date $t$ we generate a realization from the distribution of $\tilde{U}$. We obtain the corresponding observation date by replacing the random variable $\tilde{U}$ in $\eqref{eq:simulReportingDate}$ by this sampled value.



\paragraph{Step 3. Truncation} With steps 1 and 2 we have simulated an observation date for each occurred event. We split this data set into observed and hidden events. We use the data set with observed events to calibrate the model and to predict the number of hidden events. The hidden events are kept only for evaluating the prediction accuracy.









\section{A standard distribution for the time changed observation delay} \label{section:AdaptingExponential}
Modeling the time-changed observation delay with an exponential distribution has significant computational benefits. Therefore, this section puts focus on the use of the exponential distribution as a standard distribution for modeling the time-changed observation delay $\tilde{U}$. Since the exponential distribution is light-tailed it is less suited for long or heavy-tailed delays. We outline a strategy for addressing this weakness of the exponential distribution. 

Our strategy bins the possible observation delays $(s-t = 0, 1, \ldots)$ and categorizes these bins with a delay covariate $x_{s-t}^{\texttt{delay}}$. This covariate is then included in the observation \weight{} specification. For each bin we estimate a parameter to capture its effect on observation exposure. These parameters can strongly reshape the distribution, hereby overcoming many of the disadvantages of the exponential distribution. We present a maximum likelihood driven binning strategy in Appendix~\ref{section:binDelay} and then Appendix~\ref{appendix:KM} derives the same bins by linking our approach to the non-parametric Kaplan-Meier estimator \citep{KaplanMeier}. 


\subsection{Binning observation delay} \label{section:binDelay}
Our binning strategy maximizes the loglikelihood in~\eqref{eq:loglikelihoodExp} when the observation exposures depend only on the time elapsed since the event occurred, i.e.~
\begin{equation*}
	\alpha_{t, s} = \exp(\vectorF{\gamma}^{\texttt{delay}} \cdot x_{s-t}^{\texttt{delay}}) = \exp(\gamma{^\texttt{s-t}}),
\end{equation*}
where we estimate for each delay $s-t$ a separate parameter $\gamma{^\texttt{s-t}}$. Furthermore we neglect the last term in~\eqref{eq:loglikelihoodExp}, capturing the effect of the right truncation. Under these restrictions, the loglikelihood to optimize is
 \begin{align*}
 	\ell(\vectorF{\gamma} ; \vectorF{\chi} ) &= -\sum_{t=1}^\tau \sum_{v=t}^{\tau-1} \left( \sum_{s=v+1}^\tau N_{t, s}\right) \cdot \exp(\gamma^{\texttt{v-t}}) + \sum_{t=1}^\tau \sum_{s=t}^\tau N_{t, s} \cdot \log \left( 1 - \exp(- \exp(\gamma^{\texttt{s-t}}) \right) 
 \end{align*}
We compute the derivatives of $\ell(\vectorF{\gamma} ; \vectorF{\chi})$ with respect to the observation exposure parameter $\gamma^{\texttt{d}}$ for positive delays $d \in \mathbb{N}$
\begin{equation*}
	\frac{\partial \ell(\vectorF{\gamma} ; \vectorF{\chi} )}{\partial \gamma^{\texttt{d}}} = -\exp(\gamma^{\texttt{d}}) \cdot \sum_{t=1}^{\tau - d-  1} \sum_{s = t + d + 1}^\tau N_{t, s} + \frac{\exp(\gamma^{\texttt{d}})}{\exp(\exp(\gamma^{\texttt{d}})) - 1} \cdot \sum_{t=1}^{\tau - d} N_{t, t + d}.
\end{equation*}
Both sums in this expression have a logical interpretation. The first sum ($\sum_{t=1}^{\tau - 1 - d} \sum_{s = d + t + 1}^\tau N_{t, s}$) counts the number of observed events with a delay longer than $d$ days, whereas the second sum ($\sum_{t=1}^{\tau - d} N_{t, t + d}$) counts all events with a delay of exactly $d$ days. These derivatives are zero when
\begin{equation}
	\exp(\gamma^{\texttt{d}}) = -\log\left( 1 - \frac{|\texttt{delay = d}|}{|\texttt{delay $\geq$ d}|} \right), \label{eq:breakpointChoiceExponential}
\end{equation}
where $|\texttt{delay = d}|$ denotes the number of events observed with a delay of $d$ days and $|\texttt{delay > d}|$ the number of events with a delay of more than $d$ days. 

\begin{figure}[ht!]
	\center
	\includegraphics[width = 0.95\textwidth]{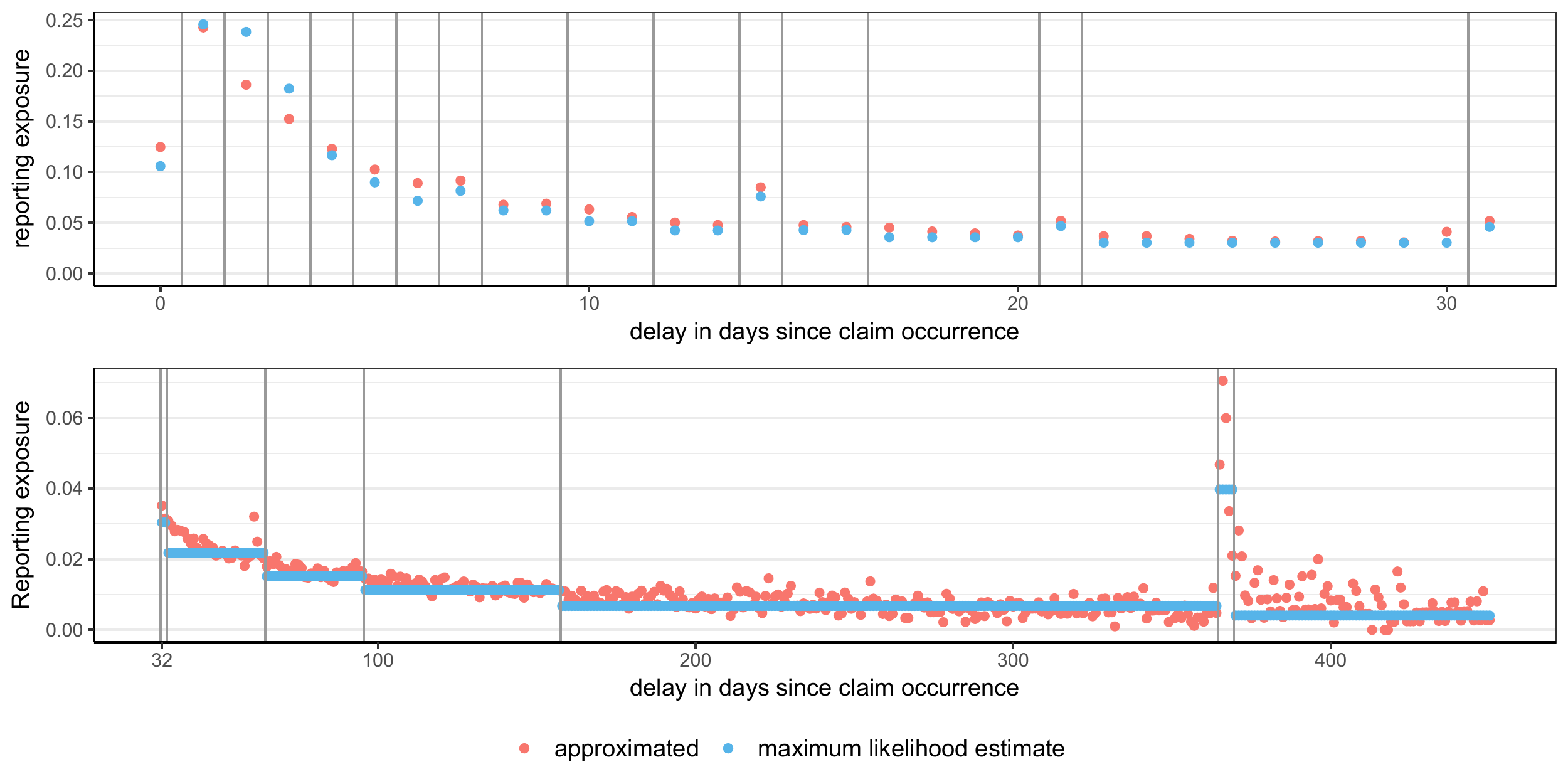}
	\caption{Observation exposure estimates for the delay effect during the first month after the accident occurrence (top) and longer delays (bottom). In red, we show estimates obtained for each delay using \eqref{eq:breakpointChoiceExponential}. The vertical lines indicate the chosen bins. Maximum likelihood estimates for the observation delay parameter corresponding to each bin in the regression structure proposed in Section~\ref{section:modelliability} are plotted in blue.}
	\label{figure:approxTheta}
\end{figure}

We propose to bin the observation delay by grouping delays for which $\eqref{eq:breakpointChoiceExponential}$ is approximately constant. Figure~\ref{figure:approxTheta} visualizes this approach for the liability insurance data set discussed in Section~\ref{section:caseStudy}. This figure shows in red the estimated delay parameters using approximation \eqref{eq:breakpointChoiceExponential}. The top panel shows the estimates for delays up to 31 days, whereas the parameters for larger delays (up to 400 days) are shown in the bottom panel. Based on this knowledge observation delay is grouped in 23 bins, separated by vertical gray bars in Figure~\ref{figure:approxTheta}. We use more bins for short delays, since for these delays $\eqref{eq:breakpointChoiceExponential}$ differs strongly. Moreover, many accidents have a short observation delay, which makes these first delays more important. As expected, this binning strategy identifies an increase in observation probability after exactly one year. In Section~\ref{section:caseStudy} we structure these bins in a categorical delay covariate $x_{s-t}^{\texttt{delay}}$ and estimate observation delay in a maximum likelihood framework. In Figure~\ref{figure:approxTheta} the fitted parameters are plotted in blue. These parameters deviate from those found using approximation \eqref{eq:breakpointChoiceExponential}, since other covariate effects were estimated simultaneously. However, the maximum likelihood estimates are close to the approximate values which makes this approximation suitable for choosing initial values in the calibration.


\subsection{A link with the Kaplan-Meier estimator} \label{appendix:KM}
We show that under the binning strategy of Appendix~\ref{section:binDelay} the time changed model has the same flexibility as the Kaplan-Meier estimator and is as such suitable for modelling a wide range of portfolios. 


The Kaplan-Meier estimator for the survival function of the observation delay random variable is
\begin{equation}
	\widehat{P(\texttt{delay} > d)} = \prod_{i=0}^{d} \left( 1 - \frac{|\texttt{delay} = i|}{|\texttt{delay} \geq i|} \right), \label{eq:hazard1}
\end{equation}
When we model the time-changed observation delay distribution $\tilde{U}$ using an exponential distribution then the survival probability for an event from occurrence day $t$ is
\begin{align}
	P \left( \texttt{delay} > d \, \mid \, \texttt{occ.~day} = t \right) 	&= P \left( \tilde{U} \geq \varphi_t(d+1) \right) \label{eq:hazard2}\\
								&= 1 - F_{\tilde{U}}\left(\sum_{i=1}^{d+1} \alpha_{t, t+i-1}\right) \nonumber\\ 
								&= \prod_{i=0}^{d} \exp \left( -\alpha_{t, t+i} \right). \nonumber 
\end{align}
Notice the similarity between this expression and the Kaplan-Meier estimator in $\eqref{eq:hazard1}$. When the observation exposure only depends on the time passed since the occurrence of the event, i.e. $\alpha_{t, t+i} := \alpha_i$, then 
\begin{align*}
	P \left( \texttt{delay} > d \right) = \prod_{i=0}^{d} \exp \left( -\alpha_{i} \right), 
\end{align*}
where $\alpha_{i}$ is the observation exposure at delay $i$. This expression no longer depends on the occurrence date $t$ of the event. The Kaplan-Meier estimator is retrieved when 
\begin{align}
	\alpha_i = -\log \left( 1 - \frac{|\texttt{delay} = i|}{|\texttt{delay} \geq i|} \right). \label{eq:structureDelay}
\end{align}
Since $\alpha_i = \exp(\gamma^{i})$, this is the same estimator we found in \eqref{eq:breakpointChoiceExponential} through maximum likelihood estimation. This show that by estimating a separate delay parameter for each delay $(d = 0, 1, \ldots)$ we obtain a model with the same flexibility as the non-parametric Kaplan-Meier estimator.

\end{document}